\documentclass[prd,preprint,aps,amsmath,amssymb,showpacs,superscriptaddress,nofootinbib]{revtex4-2}

\usepackage{graphicx} 
\usepackage{dcolumn}
\usepackage{bm}
\usepackage{hyperref}
\usepackage{color}
\usepackage[normalem]{ulem}
\usepackage{comment}
\usepackage{diagbox}

\newcommand{\MeV}{\rm MeV}

\newcommand{\dneff}{\Delta N_{\rm eff}}

\usepackage{ulem}

\begin{document}

\preprint{PITT-PACC-2119}
\preprint{MI-HET-815}

\title{Gravitational Waves from Nnaturalness}

\author{Brian Batell}
\email{batell@pitt.edu}
\affiliation{Pittsburgh Particle Physics, Astrophysics, and Cosmology Center,\\Department of Physics and Astronomy, University of Pittsburgh, Pittsburgh, USA}
\author{Akshay Ghalsasi}
\email{akg53@pitt.edu}
\affiliation{Pittsburgh Particle Physics, Astrophysics, and Cosmology Center,\\Department of Physics and Astronomy, University of Pittsburgh, Pittsburgh, USA}
\author{Matthew Low}
\email{mal431@pitt.edu}
\affiliation{Pittsburgh Particle Physics, Astrophysics, and Cosmology Center,\\Department of Physics and Astronomy, University of Pittsburgh, Pittsburgh, USA}
\author{Mudit Rai}
\email{muditrai@tamu.edu}
\affiliation{Pittsburgh Particle Physics, Astrophysics, and Cosmology Center,\\Department of Physics and Astronomy, University of Pittsburgh, Pittsburgh, USA}
\affiliation{Mitchell Institute for Fundamental Physics and Astronomy,\\Department of Physics and Astronomy, Texas A\&M
University, College Station, USA}

\date{\today}

\begin{abstract}
We study the prospects for probing the Nnaturalness solution to the electroweak hierarchy problem with future gravitational wave observatories.
Nnaturalness, in its simplest incarnation, predicts $N$ copies of the Standard Model with varying Higgs mass parameters. 
We show that in certain parameter regions the scalar reheaton transfers a substantial energy density to the sector with the smallest positive Higgs squared mass while remaining consistent with bounds on additional effective relativistic species. 
In this sector, all six quarks are much lighter than the corresponding QCD confinement scale, 
allowing for the possibility of a first-order chiral symmetry-breaking phase transition and an associated stochastic gravitational wave signal.
We consider several scenarios characterizing the strongly-coupled phase transition dynamics
and estimate the gravitational wave spectrum for each. 
Pulsar timing arrays (SKA), spaced-based interferometers (BBO, Ultimate-DECIGO, $\mu$Ares, asteroid ranging), and astrometric measurements (THEIA) all have the potential to explore new regions of  Nnaturalness parameter space, complementing probes from next generation cosmic microwave background radiation experiments. 
\end{abstract}

\maketitle

\newpage

\section{Introduction}

The naturalness puzzle associated with the Higgs mass has for several decades inspired a vision of rich dynamics underlying the electroweak scale, possibly involving supersymmetry, new strong dynamics, or extra spatial dimensions, with a host of new states within reach of high energy colliders. However, the key lessons of the Large Hadron Collider (LHC), including the existence of a Higgs boson with properties in agreement with the Standard Model (SM) predictions and the absence thus far of new degrees of freedom at the TeV scale, have led physicists to question this traditional vision and pursue new lines of attack on the hierarchy problem, see for example Refs.~\cite{Dienes:1994np,Dienes:2001se,Dvali:2003br,Dvali:2004tma,Chacko:2005pe,Burdman:2006tz,Cheung:2014vva,Graham:2015cka,Arvanitaki:2016xds,Ibanez:2017oqr,Geller:2018xvz,Cheung:2018xnu,Craig:2019fdy,Giudice:2019iwl,Craig:2019zbn,Strumia:2020bdy,Csaki:2020zqz,Arkani-Hamed:2020yna,Giudice:2021viw,TitoDAgnolo:2021nhd,Abel:2021tyt,Csaki:2022zbc}, which also include earlier relevant literature, as well as the recent reviews~\cite{Giudice:2017pzm,Craig:2022uua}.

Among these new ideas, one particularly interesting approach, known as Nnaturalness~\cite{Arkani-Hamed:2016rle}, posits $N$ mutually non-interacting copies of the SM with Higgs mass parameters distributed over the range of the cutoff of the theory.  In this way, some sectors will accidentally have Higgs mass parameters that are parametrically smaller than the cutoff, and our SM is identified with the sector having the smallest (negative) squared Higgs mass. Finally, a light `reheaton' with universal portal couplings to each sector will naturally transfer most of its energy to our sector and only slight fractional energy densities to the other sectors, allowing for a viable cosmology with small, but potentially testable, departures from $\Lambda$CDM.

By construction, experimental and observational tests of Nnaturalness are scarce, with the most robust probes coming from cosmology~\cite{Arkani-Hamed:2016rle}. The extra energy deposited in the other sectors leads to dark radiation, which can be probed in current and future generation CMB experiments~\cite{Planck:2018vyg,CMB-S4:2016ple,Bansal}.  Also, the slightly heavier neutrinos from other sectors may free stream around matter-radiation equality, which can suppress the matter power spectrum to a level that is potentially measurable~\cite{Hildebrandt:2018yau,EuclidTheoryWorkingGroup:2012gxx}.  On the other hand, the possibility of probing the additional sectors or the reheaton at accelerator experiments is remote. 
It is therefore of great interest to find additional probes of the scenario. 

In this work we investigate the prospects for probing Nnaturalness through gravitational wave (GW) signatures. The basic idea we will explore concerns the dynamics of QCD in the {\it exotic} sectors having positive squared Higgs masses.  In such exotic sectors, all six quarks are light in comparison to the corresponding QCD confinement scale. Therefore, these exotic sectors may undergo a first-order phase transition (FOPT) associated with the breaking of the corresponding $SU(6)_L \times SU(6)_R\times U(1)_A$ chiral symmetry~\cite{Pisarski:1983ms},
which in turn generates a stochastic GW signal~\cite{Witten:1984rs,Hogan:1986qda,Kosowsky:1991ua,Kosowsky:1992rz,Kosowsky:1992vn,Kamionkowski:1993fg,Kosowsky:2001xp,Dolgov:2002ra,Caprini:2007xq,Huber:2008hg,Caprini:2009yp,Hindmarsh:2013xza,Hindmarsh:2015qta,caprini2016science,Hindmarsh:2017gnf,caprini2018cosmological}. This possibility was discussed previously in Ref.~\cite{Archer-Smith:2019gzq} which, however, concluded that for a particular point in parameter space the energy density in the exotic sectors was too small to lead to a detectable GW signal.  
We revisit this possibility and identify regions of the Nnaturalness parameter space where the exotic sector with the lightest Higgs contains enough energy density to yield a detectable GW signal, yet is still consistent with current bounds on additional relativistic degrees of freedom. Depending on the details of the QCD phase transition in this sector, which are unfortunately obscure due to strong dynamics, the corresponding GW wave signal is predicted to lie in nHz $-$ Hz frequency range, with an amplitude that is potentially detectable by several current and planned GW observatories. 

The discovery  of GWs by the LIGO-Virgo-KAGRA collaborations~\cite{LIGOScientific:2016aoc,LIGOScientific:2018mvr,LIGOScientific:2020ibl,LIGOScientific:2021djp} has opened a new observational window 
to the universe in the ${\rm Hz} - {\rm kHz}$ frequency range.  
Existing and new observatories planned in the next decade and beyond will be capable of measuring GWs over a much wider frequency range and with significantly smaller amplitudes. 
While the GWs observed by LIGO-Virgo-KAGRA are sourced by mergers of compact objects such as $\mathcal{O}(1-100 \, M_{\odot})$ black holes, new observatories offer the promise of  
probing a variety of GW signals and sources. This includes stochastic GWs, 
which may be generated by FOPTs,
as well as other exotic  sources~\cite{turner1990relic,chang1998studies,Schwaller:2015tja,saikawa2017review,Blasi:2020mfx,Ellis:2020ena,Liu:2020mru, Nakai:2020oit,Vaskonen:2020lbd,DeLuca:2020agl,Ratzinger:2020koh,Namba:2020kij,Vagnozzi:2020gtf,Li:2020cjj,Co:2021lkc,Borah:2021ftr,Gouttenoire:2021jhk,Jinno:2022fom,Berger:2023pon}. 
Recently, the NANOGrav pulsar timing array (PTA)
reported evidence for a signal in the ${\cal O}(1-10)$ nHz frequency range with a Hellings–Downs correlation~\cite{Hellings:1983fr} that is characteristic of stochastic GWs in their 15-yr survey~\cite{NANOGrav:2023gor}. This result is supported by the marginal evidence for a stochastic GW signal from the European PTA~\cite{Antoniadis:2023ott} and is also consistent with results from the Parkes PTA~\cite{Reardon:2023gzh} and China PTA~\cite{Xu:2023wog}. 
This milestone, if eventually confirmed with future data, marks a new era of exploration of novel astrophysical and cosmological GW sources.

We explore the capability of several existing and future GW experiments, including PTAs, spaced-based interferometers, and astrometric measurements, to probe  
the Nnaturalness exotic sector QCD phase transition. To this end, we first identify the most promising regions of Nnaturalness parameter space for a GW signal where the first exotic sector contains a substantial energy density while remaining consistent with constraints from the CMB on new effective relativistic degrees of freedom. 
We consider several scenarios characterizing the dynamics of the exotic sector QCD phase transition, which allow us to explore a range of possible GW signals in a model-agnostic fashion.
Depending on these assumptions, we find that experiments such as the Square Kilometer Array (SKA) PTA~\cite{Janssen:2014dka}, the spaced-based interferometers LISA~\cite{2017arXiv170200786A}, BBO~\cite{Crowder:2005nr}, Ultimate-DECIGO~\cite{Braglia:2021fxn}, and $\mu$Ares~\cite{Sesana:2019vho}, and asteroid laser ranging~\cite{Fedderke:2021kuy}, and future astrometric measurements \cite{Moore:2017ity} by the proposed THEIA experiment~\cite{2018FrASS...5...11V}  have the potential to probe Nnaturalness. 
On the other hand, we find that Nnaturalness is unlikely to account for all of the stochastic GW signal recently reported by NANOGrav due to stringent CMB constraints on new relativistic degrees of freedom.

The rest of this paper is organized as follows: In Section~\ref{sec:Nnaturalness} we review the minimal Nnaturalness model with a scalar reheaton, focusing on the salient features, particularly of the first exotic sector, that will be used in our subsequent analysis of the cosmology and GW signal.  
Next, in Section~\ref{sec:cosmology}, we discuss the cosmology of the scenario and estimate the contribution from the other sectors to the extra effective relativistic degrees of freedom at late time. 
In Section~\ref{sec:GW} we discuss our estimate of the GW signal under different assumptions regarding the nature of the exotic sector QCD phase transition. Our main results are presented in Section~\ref{sec:res}, which include a delineation of the Nnaturalness parameter space that may potentially be probed by future GW observatories. Our conclusions and outlook are presented in Section~\ref{sec:con}. Appendices~\ref{sec:nn} and~\ref{sec:gstar}
 contain technical details on the reheaton decays and example estimates of the effective relativistic degrees of freedom in the different sectors at several stages of the cosmological history, respectively. 

\section{Nnaturalness}
\label{sec:Nnaturalness}

The minimal\,\footnote{Nnaturalness is more general than the minimal model considered here and only requires that the SM is not atypical among all sectors.} Nnaturalness model contains $N$ copies of the SM that are mutually decoupled. The Higgs squared mass parameters are assumed to vary uniformly from one sector to another according to the relation 
\begin{align}
\label{eq:mHi}
m^{2}_{H_i} = -\frac{\Lambda_H^{2}}{N}\left(2 i +r\right), 
\quad\quad\quad 
-\frac{N}{2} \leq i \leq \frac{N}{2}.
\end{align}
Here $i$ labels the sector, with our SM identified with the $i = 0$ sector such that $m^{2}_{H_0} = m^{2}_{H} = -(88~{\rm GeV})^2$.
The parameter $r$ controls the relative distance of $m^{2}_{H}$ from zero, with $0 \leq r < 2$, and $\Lambda_H$ is the cutoff of the theory. The {\it SM-like} sectors with $i > 0$ have negative squared mass parameters, while the {\it exotic} sectors with $i < 0$ have positive squared mass parameters. 

Besides the $N$ sectors, the other crucial ingredient in Nnaturalness is the reheaton 
which is assumed to dominate the energy density of the universe at some time following inflation. 
Ref.~\cite{Arkani-Hamed:2016rle} considered models with a scalar reheaton and models with a fermionic reheaton.  For concreteness, in this work we focus on the real scalar $\phi$ reheaton, with Lagrangian
\begin{align}
{\cal L}_\phi \supset - a \phi \sum_i |H_i|^2 - \frac{1}{2}m_\phi^2 \phi^2,
\label{eq:phi-couplings}
\end{align}
where $m_\phi$ is the reheaton mass and $a$ is a universal dimensionful coupling of the reheaton to the Higgs fields. 
The couplings in Eq.~\eqref{eq:phi-couplings} cause the reheaton to decay into all sectors. 
As we will discuss in detail below, a light reheaton, with a mass near the electroweak scale, can dominantly decay to and populate the $i = 0$ SM sector, thereby allowing for a viable cosmology and a solution to the hierarchy problem. 
Besides the variation in their Higgs mass parameters, the sectors are assumed to be identical in all respects, which implies the theory has a softly broken sector permutation symmetry.\footnote{For consistency in the large $N$ limit, it is necessary to consider an arbitrary sign for the coupling $a$ in each sector and require that $|a| \lesssim \Lambda_H/N$.  The scaling of $1/N$ ensures the loop-induced mass for the reheaton is controlled while the arbitrary sign maintains control over the loop-induced tadpole for the reheaton.} 

The SM-like sectors, due to their large, negative Higgs squared masses, undergo electroweak symmetry breaking in the familiar way, $\langle H_i \rangle \neq 0$, with the Higgs fields obtaining large vacuum expectation values (VEVs) $v_{i}^2  = -m_{H_i}^2/\lambda = v^2 (2i/r+1)$, with $\lambda$ the universal Higgs quartic coupling and $v= 246$ GeV the SM Higgs VEV. Instead, in the exotic sectors the Higgs squared masses are positive, and electroweak symmetry breaking is triggered by QCD strong dynamics through the formation of a quark condensate $\langle \overline q q \rangle_i \neq 0$. The exotic sector quarks receive masses of order $m_{q_{i}} \sim y_q y_t \Lambda_{{\rm QCD}_i}^3/m_{H_i}^2$ and therefore are all much lighter than corresponding confinement scale
$\Lambda_{{\rm QCD}_i} \sim {\cal O}(100 \, \rm MeV)$.  
Hence, these exotic sectors, with six light quark flavors, may undergo FOPTs associated with 
the breaking of the corresponding $SU(6)_L \times SU(6)_R\times U(1)_A$ chiral symmetry~\cite{Pisarski:1983ms}, which then generates a stochastic GW signal. 
We note that there is still some debate in the literature on the order of this phase transition, and we comment further on this in Sec.~\ref{sec:cosmology}. 
Assuming the phase transition is first order, the detectability of the GW signal depends on how much energy density is contained in the exotic sectors, and only the first exotic sector ($ i = -1$) may have a substantial energy density in the cosmologically allowed regions of parameter space. 
To understand this, we must carefully examine the cosmological evolution of the model, which, in any case, is of central importance in the Nnaturaless solution to the hierarchy problem. This will be discussed in detail in the Section~\ref{sec:cosmology}.

The minimal Nnaturalness model considered here is thus characterized by four parameters: $N$, $r$, $m_\phi$, and $a$.  The GW signal will be relatively insensitive to the value of $N$ because the signal originates solely from the $i=-1$ exotic sector. 
 For concreteness, for the rest of this work we fix $N = 10^4$ which allows for a solution to the little hierarchy problem, with $\Lambda_H \sim 10$ TeV, and evades any potential issues with overclosure from massive stable states from the other sectors. 
Likewise, the Nnaturalness mechanism and the GW signal are not sensitive to the precise value of the universal coupling $a$ since it cancels out in the reheaton decay branching ratios. The only requirement is that $a$ is small enough so that the reheating temperature is below the electroweak scale, which can always be satisfied. The GW signals will therefore be controlled by $m_\phi$ and $r$.

\subsection{Reheaton decays}

The fraction of the reheaton energy density transferred to each sector is proportional to the reheaton partial decay width into each sector, which we denote by $\Gamma_i$.  If the reheaton is light, with mass of order the electroweak scale, it can dominantly decay to the SM sector. Depending on the values of $m_\phi$ and $r$, there may also be a significant energy density stored in the other sectors, leading to constraints and probes from additional relativistic degrees of freedom and GWs, to be discussed in the next sections. Some details related to the reheaton decays to the SM-like and exotic sectors are provided in Appendix~\ref{sec:nn}; we now summarize their basic properties.  

In the SM and SM-like sectors electroweak symmetry breaking causes $\phi$ to mix with the corresponding physical Higgs boson $h_i$, with mixing angle $\theta_{i} \simeq a \, v_{i} / m^{2}_{h_i} \approx  a / m_{h_i}$, where $m_{h_i}$ is the physical Higgs mass for the sector. 
Thus, the $\phi$ partial decay widths to SM-like sectors scale as $\Gamma_{i} \propto 1/m^{2}_{h_i}$, with decays to the SM being the largest. 
Eq.~\eqref{eq:mHi} implies that $m_{h_i}$ decreases as $r$ is increased. Thus, the fractional energy densities deposited in the $i \geq 1$ SM-like sectors tend to increase as $r$ increases. 

In the exotic sectors, the small effects of electroweak symmetry breaking from QCD can be neglected as far as the decays of the reheaton are concerned. 
Except for perhaps the lightest exotic sectors, we expect $m_\phi < m_{H_{i}}$, in which case the reheaton decays to the exotic sector $i$ mainly proceed through a loop via $\phi \rightarrow W_i W_i, B_i B_i$, with a decay width given by $\Gamma \propto 1/m^{4}_{H_i}$, or through a four-body decay $\phi \rightarrow H_i^*  H_i^{*}$. Therefore, the energy density stored in the heavier exotic sectors is generally insignificant in the viable regions of parameter space. Only the first exotic sector ($i = -1$) may potentially receive a substantial portion of the reheaton's energy density.  As can be seen from Eq.~\eqref{eq:mHi}, as $r$ is increased the first exotic sector Higgs mass $m_{H_{-1}}$ decreases, and for $m_{H_{-1}} \sim m_\phi/2$ or below the reheaton may have a sizable or even dominant branching ratio into the lightest exotic sector via the two- or three-body decay $\phi \rightarrow H_{-1}  H_{-1}^{(*)}$. 

Besides the general trends outlined above, when the reheaton mass is close to the SM Higgs mass, $m_\phi \sim m_h$, there is a resonant enhancement in $\phi-h$ mixing, $\theta_{\rm SM} \simeq a v/(m_h^2 - m_\phi^2)$, which enhances the decays of the reheaton to the SM sector. Another such enhancement occurs for $m_\phi \gtrsim 2 m_W$, $2 m_Z$, when reheaton decays to on-shell SM weak bosons open up. 

\subsection{Properties of the First Exotic Sector}

As explained above, among the exotic sectors only the first one ($i = -1$) may have a significant fraction of the reheaton energy density in the cosmologically viable regions of parameter space. Thus, it is only this sector which may furnish a potentially detectable  stochastic GW signal from its corresponding QCD FOPT. 

In this sector, the Higgs squared mass is positive. Hence, in the absence of QCD strong dynamics, electroweak symmetry would not be spontaneously broken, and all fermions and gauge bosons would be massless.  However, as in the SM sector, QCD in this sector becomes strongly interacting at scales of order 1 GeV, and a quark condensate forms, $\langle \bar q q \rangle_{-1} \sim 4 \pi f_{\pi_{-1}}^3$ with $f_{\pi_{-1}}$ the corresponding pion decay constant, spontaneously breaking the approximate global chiral symmetry $SU(6)_L \times SU(6)_R  \rightarrow SU(6)_V$. This condensate thus also breaks the weakly gauged electroweak subgroup down to electromagnetism in the usual way. 
Of the 35 pions associated with this chiral symmetry breaking, three linear combinations form the true Nambu-Goldstone bosons eaten by the $W^{\pm}$  and $Z$ bosons to give them masses. 
The quark condensate triggers an effective tadpole for the Higgs field, inducing a VEV $\langle H_{-1} \rangle \neq 0$.
This in turn generates masses for the leptons and quarks (though the latter are confined into hadrons at low energy). The chiral symmetry is explicitly broken by the Yukawa couplings and the electroweak gauge interactions, and this explicit breaking will cause the remaining 32 pions to obtain masses, i.e., they are pseudo-Nambu Goldstone bosons (pNGBs).

We now provide some results for the mass spectrum of the exotic sector states lighter than confinement scale, which will be relevant in our discussion of the cosmology and GW signal. For our quoted numerical estimates in the following, we choose a benchmark $m_{H_{-1}} = 70$ GeV and take $f_{\pi_{-1}} = 30$ MeV.\footnote{We assume that $f_{\pi_{i}}$ scales linearly with $\Lambda_{{\rm QCD}_i}$.  The value of $\Lambda_{{\rm QCD}_i}$ changes weakly with $i$ due to the change in the quark mass thresholds. In particular, taking $\alpha_S(m_Z)_{\rm SM} \simeq 0.118$ as input and using one-loop running, we find that  $\Lambda_{{\rm QCD}_{-1}}/\Lambda_{{\rm QCD}_{\rm SM}} \approx 0.3$.}
The electroweak gauge boson masses are given by $m_{W_{-1}} = (\sqrt{3}/2) g f_{\pi_{-1}}$ and  $m_{Z_{-1}} = (\sqrt{3}/2) \sqrt{g^2+g'^2} f_{\pi_{-1}}$, resulting in corresponding numerical estimates of order 20 MeV. 
The 32 pions receive masses of parametric size $m_{\pi_{-1}} \sim  4 \pi   \sqrt{y_q \,y_t} \, f_{\pi_{-1}}^2/m_{H_{-1}}$ from the explicit chiral symmetry breaking by the quark Yukawa couplings. 
We find the pions range in mass between about 1 keV and 100 keV. The leptons masses are 
$m_{\ell_{-1}}  \sim 4 \pi  \, y_\ell \,  y_t \, f_{\pi_{-1}}^3/(2 \, m_{H_{-1}}^2)$ yielding estimates of $0.1 \,{\rm meV},  20 \,{\rm meV},  0.3 \,{\rm eV}$ for the electron, muon, and tau, respectively. Neutrinos are expected to be extremely light, $m_{\nu_{-1}} < 10^{-11}$~eV, while the photon is massless.

\section{Cosmological Evolution}
\label{sec:cosmology}

The cosmological history of the Nnaturalness model starts when the reheaton dominates the energy density of the universe. 
The reheaton then decays into all available channels, reheating the universe such that each sector is populated with an energy density that scales with the reheaton's partial decay width in that sector $\rho_i/\rho_{\rm SM} \simeq \Gamma_i/\Gamma_{\rm SM}$.  Each sector thermalizes within its own sector with corresponding energy and entropy densities  given by 
\begin{equation}
\rho_i = \frac{\pi^2}{30}\, g_{*\rho,i} \, \xi_i^4 \, T^4,
\quad\quad\quad
s_i = \frac{2\pi^2}{45} \, g_{*s,i} \, \xi_i^3 \, T^3,
\end{equation}
where $T$ denotes the SM temperature, $\xi_i \equiv T_i/T$ is the ratio of $i$th sector temperature to that of the SM, and $g_{*\rho,i}$ ($g_{*s,i}$) denotes the the effective number of relativistic (entropy) degrees of freedom in sector $i$. We refer the reader to Appendix~\ref{sec:gstar} for example estimates of the effective relativistic degrees of freedom at various cosmological epochs. 
A lower energy density equates to a colder temperature (relative to the temperature of the SM bath) because $\rho_i \propto T_i^4$.  
Thus sectors with larger $|i|$ will be increasingly cold.\footnote{We assume that the baryon asymmetry in all SM-like and all exotic sectors is negligible otherwise the additional matter would overclose the universe~\cite{Arkani-Hamed:2016rle}.  
}

The reheat temperature $T^{\rm RH}$ 
of the SM bath can be taken as a free parameter since it is governed by the free coupling $a$, and we consider $\Lambda_{\rm QCD_{-1}} \lesssim T^{\rm RH} \lesssim v$. The upper bound must be imposed to avoid finite temperature corrections to the Higgs potential which would spoil the Nnaturalness mechanism, while we impose the lower bound so that the first exotic sector is reheated above its corresponding confinement scale such that the sector experiences a cosmological FOPT. 
We will fix $T^{\rm RH}   = 100$ GeV for concreteness.  
The temperature ratio for sector $i$ at reheating is given by
\begin{equation}
\label{eq:xi-RH}
\xi_i^{\rm RH} = \left[  \frac{  g^{\rm RH}_{*\rho,{\rm SM}} }{  g^{\rm RH}_{*\rho,i}  } \frac{\Gamma_i}{\Gamma_{\rm SM}}    \right]^{1/4}.
\end{equation}

The SM sector then proceeds following the usual cosmological evolution, while  each of the SM-like sectors 
evolves in a similar way to the SM sector.  
In particular, the ordering of neutrino decoupling, electron-positron annihilation, and photon recombination in these sectors is the same as in the SM~\cite{Choi:2018gho}.   The radiation in each of these sectors, in the form of free-streaming photons and neutrinos, will contribute to $\Delta N_{\rm eff}$ which measures additional relativistic degrees of freedom relative to the SM.  
As discussed in the previous section, of the exotic sectors only the first one may have a substantial energy density, and the cosmological evolution of this sector features several qualitative differences from that of the SM-like sectors. We will outline these differences in detail below, but we point out here that because the spectra of particles in the first exotic sector are much lighter than in the SM sector,  the photons can be interacting until much later times.  These sectors also contribute to $\Delta N_{\rm eff}$ but  behave as interacting radiation rather than free streaming.  The dominant cosmological signal is thus an unavoidable contribution to $\Delta N_{\rm eff}$ whose size is determined by the relevant partial width into the SM sector, compared to the sum of all other sectors.\footnote{Though it will not be studied in detail here, another cosmological signal comes from the presence of many additional species of neutrinos from the SM-like sectors.  These may free stream near matter-radiation equality and suppress the matter power spectrum~\cite{Banerjee:2016suz,Bansal}}

\subsection{Cosmology of the First Exotic Sector}

Once populated by the decay of the reheaton, the first exotic sector thermalizes and cools as the universe expands. 
During this initial period of evolution, all degrees of freedom are essentially massless except for the Higgs doublet $H_{-1}$. 

As the exotic sector cools to temperatures of order $\Lambda_{\rm QCD_{-1}} \sim {\cal O}(100)$ MeV, a chiral symmetry breaking phase transition is precipitated by the formation of the quark condensate. The conventional wisdom, due to an argument of Pisarski and Wilczek \cite{Pisarski:1983ms}, is that this phase transition is first order. 
Employing a linear sigma model description of the quark bilinear order parameter, they performed a renormalization group analysis using a perturbative $\epsilon$ expansion and noted the absence of infrared stable fixed points for $N_f\geq 3$ light quark flavors, which is indicative of a first-order phase transition. 
Subsequent studies using phenomenological models have confirmed this result, see, e.g., Ref.~\cite{Butti:2003nu}. 
This question has also been studied at various points on the lattice over the past decades, with some confirming the claim of a first-order transition for $N_f\geq 3$~\cite{Iwasaki:1995ij,Karsch:2003jg} and others challenging it~\cite{Cuteri:2021ikv}. Thus, the question of the order of the phase transition is still an open one and further study is needed to settle the issue, see Ref.~\cite{Aarts:2023vsf} for some perspectives in this direction. We will follow the conventional wisdom and assume that the exotic sector phase transition with six light flavors is first order. 

The phase transition commences at the critical temperature  $T_{-1}^{\rm crit}$, 
the point at which the potential energy of the true and false vacua are the same. For the exotic sector QCD with six massless quarks, we take $T_{-1}^{\rm crit} = 85~\MeV$, which is estimated with order 20$\%$ uncertainty~\cite{Braun:2006jd}.  Starting from the exotic sector in the symmetric phase, bubbles of true vacuum nucleate, expand, and merge, such that eventually the sector ends up in the broken phase.  The nucleation temperature, $T_{-1}^{\rm nuc} \lesssim T_{-1}^{\rm crit}$, marks the point at which the first bubbles nucleate. After nucleation, the bubbles take time to percolate until $34\%$ end up in the true vacuum corresponding to temperature  $T_{-1}^{\rm perc}$~\cite{Ellis:2018mja}. Thus it is reasonable to assume $T_{-1}^{\rm{perc}} \lesssim 85~\MeV$, and we consider the range $50~\MeV \leq$ $T_{-1}^{\rm perc} \leq 85~\MeV$. 
The temperature of the SM at percolation is $T^{\rm perc} = T_{-1}^{\rm{perc}}/\xi_{-1}^{\rm perc} $ where $\xi_{-1}^{\rm perc}$ is the temperature ratio between the exotic sector and SM right before  percolation, i.e., in the unbroken phase.  

The strength of the exotic sector phase transition is characterized by the the parameters $\alpha_{-1}$ and $\alpha_{\rm  tot}$, which are defined as 
\begin{align}
    \label{eq:alphas}
    \alpha_{-1} &= \frac{\Delta \theta_{-1}}{\rho_{-1}^{\rm perc}} , \\
    \alpha_{\rm  tot} &= \frac{\Delta \theta_{-1}}{\rho_{\rm tot}^{\rm perc}}  =   \alpha_{-1} \frac{\rho_{-1}^{\rm perc}}{\rho_{\rm tot}^{\rm perc}},
\end{align}
where $\Delta \theta_{-1}$ is the difference in the trace of the energy momentum tensor in the unbroken and the broken phases. 
Due to the strongly-coupled dynamics during the exotic sector QCD phase transition, we will not be able to provide a first principles calculation of $\alpha_{-1}$. Instead, we consider several distinct scenarios for the phase transition dynamics with varying choices for $\alpha_{-1}$.
We defer a detailed discussion of the considerations underlying these assumptions to Sec.~\ref{sec:GW}.

Once the phase transition concludes, the exotic sector is reheated to a temperature $T_{-1}^{\rm rh}.$\footnote{We note that this reheating of the exotic sector due to the corresponding QCD phase transition (labeled by `rh') should be distinguished from the reheating when the reheaton decays (labeled by `RH').} Assuming an instantaneous transition from percolation to reheating and using energy conservation, we may write 
\begin{equation}
\label{eq:FOPT-energy-conservation}
\rho_{-1}^{\rm perc} +\Delta V_{-1} = \rho_{-1}^{\rm rh},
\end{equation}
where $\Delta V_{-1}$ is the difference in free energy. 
Using Eqs.~(\ref{eq:alphas},\ref{eq:FOPT-energy-conservation}) and assuming $\Delta V_{-1} \simeq \Delta \theta_{-1}$~(see e.g., Ref.~\cite{Bringmann:2023opz}), 
we may write 
\begin{align}
\label{eq:Trhi}
T_{-1}^{\rm rh}
=T_{-1}^{\rm perc} \,
 (1+ \alpha_{-1})^{1/4} \left[\frac{ g_{*\rho,-1}^{\rm perc}  }{ g_{*\rho,-1}^{\rm rh}  } \right]^{1/4} .
\end{align}
The QCD FOPT results in entropy production in the exotic sector, which may be encoded in the ratio of entropy densities before and after the QCD phase transition, 
\begin{equation}
\label{eq-DS-1}
D_{s,-1} \equiv \frac{s_{-1}^{\rm rh}}{s_{-1}^{\rm perc }} = \frac{ g_{*s,-1}^{\rm rh} \, (T_{-1}^{\rm rh})^3 }{ g_{*s,-1}^{\rm perc} \, (T_{-1}^{\rm perc})^3} 
=(1+ \alpha_{-1})^{3/4} \left[\frac{ g_{*s,-1}^{\rm rh} }{ g_{*s,-1}^{\rm perc} }\right] \left[\frac{ g_{*\rho,-1}^{\rm perc}   }{ g_{*\rho,-1}^{\rm rh} } \right]^{3/4},
\end{equation}
where we have used Eq.~\eqref{eq:Trhi}.
The temperature of the SM at the end of the phase transition is $T^{\rm rh} = T_{-1}^{\rm rh}/\xi_{-1}^{\rm rh} $ where $\xi_{-1}^{\rm rh}$
 is the temperature ratio between the exotic sector and SM right at the end of the phase transition, i.e., in the broken phase. Assuming instantaneous reheating we have  $T^{\rm rh} = T^{\rm perc}$. 

Following the QCD phase transition,  entropy is conserved in the exotic sector throughout its subsequent evolution. As discussed in the previous section, the light degrees of freedom with masses below $\Lambda_{\rm QCD_{-1}}$ consist of the electroweak gauge bosons, pions, charged leptons, neutrinos, and photons. As the temperature drops below their masses, the electroweak gauge bosons and pions leave the exotic sector bath.  
Interestingly, neutrinos in this sector typically decouple while both the muon and tau are relativistic. To see this, we estimate the neutrino scattering rate as 
$\Gamma_{\nu,-1} \sim G_{F_{-1}}^2(\xi_{-1} T)^5$ and compare it to the Hubble rate. Noting that $G_{F_{-1}} \sim f_{\pi_{-1}}^{-2}$ in the exotic sector, this gives the decoupling temperature as 
\begin{equation}
\label{eq:T-1nu-decouple}
T_{-1}^{\nu\, {\rm dec}} \sim \xi_{-1}\left(   \frac{f_{\pi_{-1}}^4}{M_{\rm Pl} \xi_{-1}^5   } \right)^{1/3} \sim 10~{\rm eV}  ~~~~~ {\rm for }~~~~ \xi_{-1} \sim 0.3 \, .
 \end{equation}
Given that the charged lepton masses discussed in the previous section are typically below the eV scale, we see from Eq.~\eqref{eq:T-1nu-decouple} that neutrino decoupling in the exotic sector typically happens before electrons, muons, and taus annihilate. 
Following neutrino decoupling, and somewhat before recombination, the taus annihilate and heat the photon bath relative to the neutrinos by a factor 
\begin{equation}
\label{eq:T-1nu-late}
\frac{T_{-1}^\nu }{T_{-1}} = \left( \frac{18}{25}  \right)^{1/3}   ~~~~ {\rm for}  ~~~~~ T_{-1}  < m_{\tau_{-1}},
\end{equation}
which can be derived in the standard way using entropy conservation arguments. 
Thus, near recombination the exotic sector relativistic species comprise photons, neutrinos, electrons, and muons. The muons eventually annihilate at late times while the electrons and photons remain in equilibrium until today. 

\subsection{$\Delta N_{\rm eff}$ in Nnaturalness}

The most important constraint on Nnaturalness comes from bounds on $\dneff$ during the epoch of recombination. 
Bounds from Planck, namely ${\rm Planck+Lensing+BAO}$~\cite{Planck:2015fie}, constrain free streaming 
$\Delta N_{\rm eff}^{\rm CMB} \leq 0.3$ (all bounds quoted here are at the $95\%$ confidence level).  
The exotic sector more closely corresponds to an interacting fluid which results in a slightly weaker bound of 
$\Delta N_{\rm eff}^{\rm CMB} \leq 0.45$~\cite{Schoneberg:2021qvd}.
There is a well-known tension between data from Planck and data from SH0ES, but when the data from SH0ES~\cite{Riess:2020fzl} is incorporated the bound on interacting radiation is further relaxed to 
$\Delta N_{\rm eff}^{\rm CMB} \leq 0.7$~\cite{Blinov:2020hmc}. 
We use $\Delta N_{\rm eff}^{\rm CMB} \leq 0.7$ as the default constraint on $\Delta N_{\rm eff}^{\rm CMB}$.
Comparable bounds can be placed on $\dneff$ during the epoch of big bang nucleosynthesis (BBN), however, Nnaturalness generically predicts $\Delta N_{\rm eff}^{\rm CMB} > \Delta N_{\rm eff}^{\rm BBN}$.  

We evaluate $\Delta N_{\rm eff}^{\rm CMB}$ in the Nnaturalness model near the epoch of recombination at a SM temperature $T^{\rm CMB} = 0.3$ eV. Including the contributions from all sectors, this is given by
\begin{align}
\label{eq:DNeff1}
\Delta N_{\rm eff}^{\rm CMB} = \frac{8}{7} \left( \frac{11}{4} \right)^{4/3} \sum_{i \neq 0} \left[\frac{g_{*\rho,i}^{\rm CMB}}{2} \right](\xi^{\rm CMB}_i)^4.
\end{align}
We can relate  
$\xi_i^{\rm CMB}$ 
to 
$\xi_i^{\rm RH}$ 
in Eq.~\eqref{eq:xi-RH}, which depends on the reheaton partial decay width ratio $\Gamma_i/\Gamma_{\rm SM}$ and thus on the Nnaturalness model parameters $m_\phi$ and $r$. 

We first consider the contribution to Eq.~\eqref{eq:DNeff1} from the SM-like sectors.  
Using the fact that the total entropy in both the SM sector and the SM-like sectors is conserved between the epochs of reheating and recombination, along with Eq.~\eqref{eq:xi-RH}, we  obtain
\begin{align}
\label{eq:DNeff-SM-like}
\Delta N_{{\rm eff},i>0}^{\rm CMB}  = \frac{8}{7} \left( \frac{11}{4} \right)^{4/3}
 \left[\frac{g_{*\rho,{\rm SM}}^{\rm RH}}{2} \right]  
 \left[ \frac{ g_{*s,{\rm SM}}^{\rm CMB} }{ g_{*s,{\rm SM}}^{\rm RH} }   \right]^{4/3} 
  \sum_{i > 0}  \left[\frac{ g_{*\rho,i}^{\rm CMB}   }{ g_{*\rho,i}^{\rm RH}  } \right]    \left[\frac{ g_{*s,i}^{\rm RH}   }{ g_{*s,i}^{\rm CMB} } \right]^{4/3}\,\frac{\Gamma_i}{\Gamma_{\rm SM}}.
\end{align}

For the exotic sectors, only the first such sector may potentially give a significant contribution to $\Delta N_{\rm eff}$, so we focus our discussion on that sector. In comparison to the SM-like sectors, one important difference in this sector is the entropy production due to the QCD FOPT. To account for this, we first relate  $\xi_{-1}^{\rm CMB}$ to $\xi_{-1}^{\rm rh}$
using the fact that entropy is conserved between the end of the exotic sector QCD phase transition and CMB epoch.  
Next, we account for the  change in the exotic sector temperature during the phase transition, from the time of percolation to reheating, given by Eq.~\eqref{eq:Trhi}.
This equation gives a relation between  $\xi_{-1}^{\rm rh}$ and $\xi_{-1}^{\rm perc}$. 
Finally, we may again use entropy conservation to relate the $\xi_{-1}^{\rm perc}$ to $\xi_{-1}^{\rm RH}$. The final result for the first exotic sector contribution $\Delta N_{{\rm eff},-1}^{\rm CMB}$ is
\begin{align}
\label{eq:DNeff-exotic}
\Delta N_{{\rm eff},-1}^{\rm CMB}  = \frac{8}{7} \left( \frac{11}{4} \right)^{4/3}
 \left[\frac{g_{*\rho,{\rm SM}}^{\rm RH}}{2} \right]  
 \left[ \frac{ g_{*s,{\rm SM}}^{\rm CMB} }{ g_{*s,{\rm SM}}^{\rm RH} }   \right]^{4/3} 
  \left[\frac{ g_{*\rho,-1}^{\rm CMB}  }{ g_{*\rho,-1}^{\rm RH}  } \right]  
   \left[\frac{ g_{*s,-1}^{\rm RH}   }{ g_{*s,-1}^{\rm CMB}  } \right]^{4/3} 
 D^{4/3}_{s,-1}\,
 \frac{\Gamma_{-1}}{\Gamma_{\rm SM}}.
\end{align}
In comparison to the contributions from SM-like sectors in Eq.~\eqref{eq:DNeff-SM-like}, one key difference in Eq.~\eqref{eq:DNeff-exotic} is the presence of the factor $D_{s,-1}$ given in Eq.~\eqref{eq-DS-1}, which encodes the entropy production in the exotic sector due to the FOPT.  Using the benchmark values for the relativistic degrees of freedom given in Appendix~\ref{sec:gstar}, Eq.~(\ref{eq:DNeff-exotic}) gives the relation $\Gamma_{-1}/\Gamma_{\rm SM}  \approx 0.1 (1+\alpha_{-1})^{-1}\left(\Delta N^{\rm CMB}_{\rm eff,-1}/0.7\right)$.

\begin{figure}
 \includegraphics[width=0.85\linewidth]{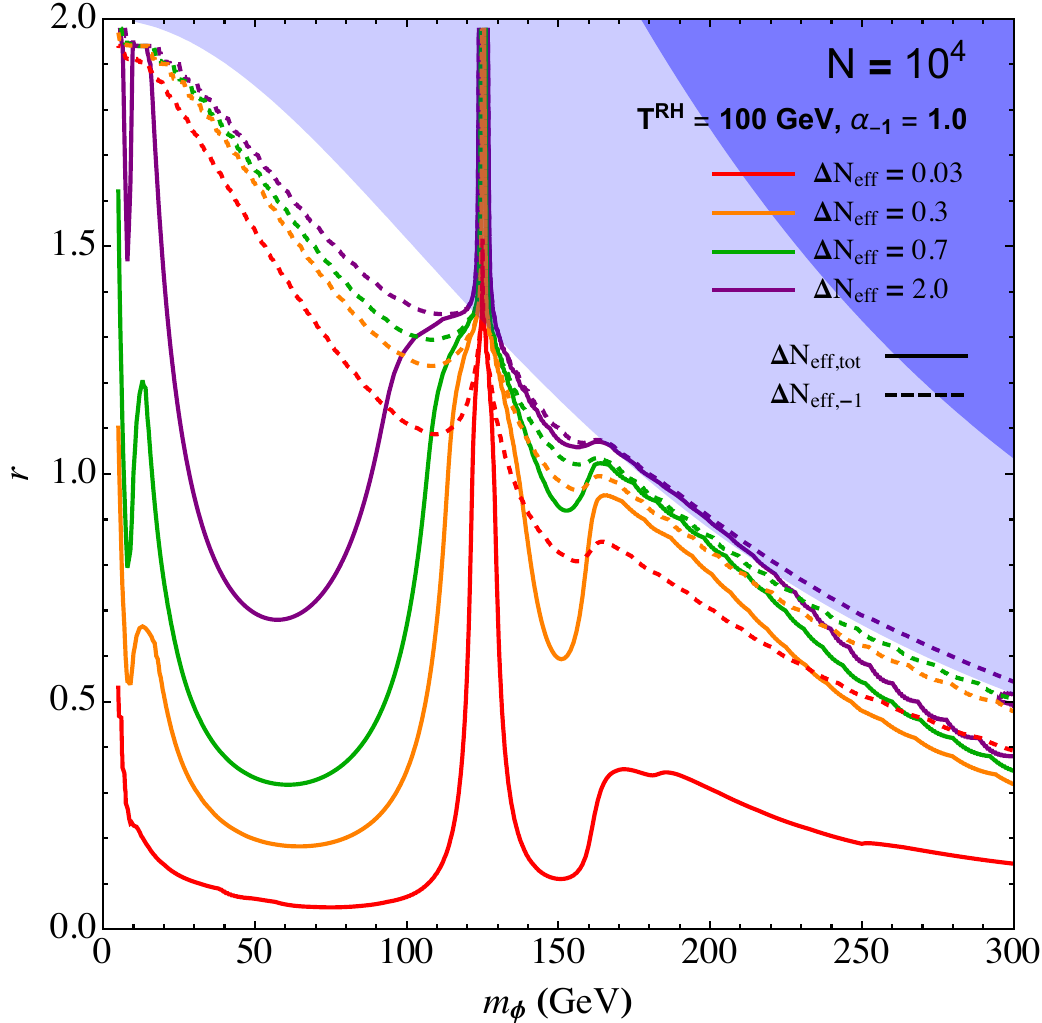}
\caption{Contours of $\Delta N_{\rm eff}$ in the $r$ vs. $m_\phi$ space, evaluated at the time of the CMB.
The number of sectors is $N=10^4$, the reheating temperature is $T^{\rm RH} = 100~{\rm GeV}$, and the phase transition strength is $\alpha_{-1} = 1$.
The solid contours show the total $\Delta N_{\rm eff}$ contribution while the dashed contours show the contribution only from the first exotic sector.  The light blue shaded region shows where the decay $\phi \to H_{-1} H_{-1}$ is on-shell and the dark blue shaded region shows where the decay $\phi \to H_{-2} H_{-2}$ is on-shell.} 
\label{fig:mPhi_r_parameterSpace}
\end{figure}

Using Eqs.~(\ref{eq:DNeff-SM-like},\ref{eq:DNeff-exotic}), in Fig.~\ref{fig:mPhi_r_parameterSpace} we show several contours of 
$\Delta N_{\rm eff}^{\rm CMB}$ 
in the parameter space of $m_\phi$ and $r$.  The solid contours show the total 
$\Delta N_{\rm eff}^{\rm CMB}$  
from all sectors.  We see that for $r \lesssim 0.2$ any mass of $\phi$ passes constraints from 
$\Delta N_{\rm eff}^{\rm CMB}$.
For larger values of $r$ there are two viable regions.  The first region is where $110~{\rm GeV} \lesssim m_\phi \lesssim 140~{\rm GeV}$.  Here the mixing between the reheaton and the SM Higgs grows much larger than the mixings between the reheaton and the Higgs particles from the other sectors.  The large relative energy density in the SM means 
$\Delta N_{\rm eff}$ is small.  Values of $r \gtrsim 1$ are possible in this region.
The second region, which permits $r \gtrsim 0.5$ is where $160~{\rm GeV} \lesssim m_\phi \lesssim 230~{\rm GeV}$.  In this region the decay of the reheaton to a pair of SM $W$ bosons goes on-shell which increases the relative energy density in the SM.  As the mass of the reheaton increases the energy density in the SM-like 
sectors and exotic sectors grows which leads to the upper limit of this region.  

 The dashed contours show $\Delta N_{{\rm eff},-1}^{\rm CMB}$ and demonstrate that there is even viable space where the energy density in the $i=-1$ exotic sector is larger than the sum over all of the SM-like sectors.  The primary reason this is possible is illustrated by the light blue shaded region which shows where the two-body decay $\phi \to H_{-1} H_{-1}$ 
goes on-shell increasing its branching ratio substantially.  It is this region where 
$\Delta N_{\rm eff}$ 
is as large as possible, while not violating existing constraints, and is dominated by the $i=-1$ exotic sector that a GW signal may be observable.

\section{Gravitational Wave Signal}
\label{sec:GW}

The first exotic sector with positive Higgs squared mass is predicted to have a substantial energy density, consistent with bounds on $\Delta N_{\rm eff}$, in certain regions of the Nnaturalness parameter space. This sector may experience a cosmological QCD FOPT, and we investigate its associated stochastic GW signal.

A cosmological FOPT can produce a stochastic GW signal due to several effects. 
During the phase transition, GWs are produced through collisions of bubble walls~\cite{Kosowsky:1992rz,Kosowsky:1992vn,Caprini:2007xq}. 
Additionally, production of GWs following the phase transition occurs due to sounds waves~\cite{Hindmarsh:2013xza,Hindmarsh:2015qta,Hindmarsh:2017gnf} and magnetohydrodynamic turbulence~\cite{Kosowsky:2001xp,Dolgov:2002ra,Caprini:2009yp} in the plasma. 
The relevant physical quantity characterizing the GW signal is the differential GW density parameter, $\Omega_{\rm GW}(f) = ({1}/{\rho_c}) \,d\rho_{\rm GW}/{d \log f}$,
where $f$ is the frequency of the GW and $\rho_c$
is the critical density. 
Sophisticated numerical simulations have been performed to properly model these dynamical processes and predict the resulting GW spectrum. 
The GW spectrum at emission can then be conveniently described by a semi-analytical parameterization that is fit to the results of numerical simulations. Following Ref.~\cite{Breitbach:2018ddu}, which makes use of the results from Refs.~\cite{Huber:2008hg,Hindmarsh:2015qta,Caprini:2009yp}, we employ the following parameterization:
\begin{equation}
\label{eq:OmegaGW-emission}
\Omega^{\rm em}_{{\rm GW}}(f_{\rm em}) =\sum_{I = {\rm BW,\,SW}} N_I\, \Delta_I(v_{\rm w}) \,\left( \frac{\kappa_I(\alpha_{-1}) \, \alpha_{\rm tot} }{1+ \alpha_{\rm tot} }  \right)^{p_I}   \left( \frac{H }{\beta }  \right)^{q_I} s_I(f_{\rm em}/f_{{\rm p},I}),
\end{equation}
where $I = {\rm BW, SW}$ denotes bubble walls and sound waves, respectively, and $f_{\rm em}$ is the GW frequency at emission. 
We do not consider the contribution from turbulence in this work since this source suffers from significant uncertainties~\cite{Brandenburg:2017neh,RoperPol:2019wvy,Brandenburg:2021tmp,Dahl:2021wyk,Auclair:2022jod}.  

We see from Eq.~\eqref{eq:OmegaGW-emission} that the GW spectrum depends on the phase transition strength parameters $\alpha_{-1}$ and $\alpha_{\rm tot}$, defined in Eq.~\eqref{eq:alphas}, the phase transition duration parameter $\beta/H$, the bubble wall velocity $v_{\rm w}$, and the efficiency factors  $\kappa_{\rm BW}$ (fraction of the vacuum energy carried by the bubble walls during collision) and $\kappa_{\rm SW}$ (energy fraction transferred to plasma bulk motion). We will return shortly to discuss our assumptions for these quantities, which depend on the detailed nature of the exotic sector QCD phase transition.
For the other quantities appearing in Eq.~\eqref{eq:OmegaGW-emission} we use the results from Refs.~\cite{Caprini:2019egz,Huber:2008hg}. For the normalization factors, we have $(N_{\rm BW},N_{\rm SW}) = (1, 0.159)$.
The velocity factor takes into account a potential suppression due to the wall velocity, with $(\Delta_{\rm BW},\Delta_{\rm SW}) = ((0.11 v^{3}_{\rm w})/(0.42 + v^{3}_{\rm w}),1)$.  
The exponents are given by $(p_{\rm BW},p_{\rm SW}) = (2,2)$ and $(q_{\rm BW},q_{\rm SW}) = (2,1)$. The spectral shape functions and corresponding peak frequencies are taken to be
\begin{align}
    \label{eq:spectrum}
    s_{\rm BW}(x) &= \frac{3.8 \, x^{2.8}}{1+2.8 \, x^{3.8}}, ~~~~~~~ s_{\rm SW}(x) =x^{3}\left(\frac{7}{4+3 \, x^{2}}\right)^{7/2} \nonumber, \\
   f_{\rm p, BW} &= 0.23\, \beta, \qquad\quad  ~~~~~~~ f_{\rm p, SW} = 0.53 \, \beta/v_{\rm w}.
\end{align}
For the sound wave contribution, we also include a additional suppression factor \cite{Ellis:2020awk,Guo:2020grp} for large $\beta/H$ given by 
\begin{equation}
\Upsilon_{\rm SW} \simeq {\rm min}\left[1,\frac{3.38 \, {\rm max}[v_{\rm w}, c_{\rm s}]}{\beta / H} \sqrt{\frac{1+\alpha_{-1}}{\kappa_{\rm  SW}\alpha_{-1}}}\,\right],
\end{equation}
where $c_{\rm s} = 1/\sqrt{3}$ is the speed of sound in the relativistic plasma.

In obtaining the observed spectrum today, one must account for the expansion of the universe from the time of GW emission until today, which redshifts both the energy density and the GW frequency:
\begin{equation}
\label{eq:OmegaGW-today}
h^2 \, \Omega_{\rm GW}^0(f) = h^2  {\cal R} \, \Omega^{\rm em}_{\rm GW}\left( \frac{a^0}{a^{\rm perc}} f \right).
\end{equation}
Here $\Omega_{\rm GW}^0$ ($\Omega^{\rm em}_{\rm GW}$) denotes the spectrum today (at emission), $f$ is the frequency today, $a^0$ ($a^{\rm perc}$) is the scale factor today (at percolation), and ${\cal R}$ is a redshift factor. 
We define the time of emission to coincide with the time of percolation, when a substantial fraction of the universe is filled with bubbles of the true vacuum.
Neglecting the small effect of entropy production during the exotic sector QCD phase transition, the relevant factors in Eq.~\eqref{eq:OmegaGW-today} are given by
\begin{align}
\frac{a^0}{a^{\rm perc}} 
 & =  \left[ \frac{g_{*s,{\rm tot}}^{\rm perc}}{g_{*s,{\rm tot}}^{\rm 0}}   \right]^{1/3} \frac{T^{\rm perc}}{T^{0}}, \nonumber \\
 h^2 {\cal R} & = h^2 \left(\frac{a^{\rm perc}}{a^0}\right)^4 \left(\frac{H^{\rm perc}}{H^0}\right)^2   = h^2 \Omega_{\gamma}^0 \left[ \frac{g_{*\rho,{\rm tot}}^{\rm perc}}{2}   \right]
 \left[ \frac{g_{*s,{\rm tot}}^{0}}{g_{*s,{\rm tot}}^{\rm perc}}   \right]^{4/3},
\end{align}
where $T^0 = 2.725\, {\rm K}  \approx  0.235 \,{\rm meV}$ is the present temperature of the CMB, $H^{0}$ ($H^{\rm perc}$) is the Hubble parameter today (at percolation) with  $H^0  = 100\, h \, {\rm km}\, {\rm Mpc}^{-1} \,{\rm s}^{-1}$ and 
$h^2\Omega_{\gamma}^0 \approx 2.47 \times 10^{-5}$ is the present photon density parameter. 
The factors counting relativistic degrees of freedom are given by
\begin{align}
    \label{eq:R}
    g_{*\rho,\rm tot}^{\rm perc} & \simeq g_{*\rho, \rm SM}^{\rm perc} + g_{*\rho,-1}^{\rm perc} (\xi_{-1}^{\rm perc})^{4} + \sum_{i>0} g_{*\rho,i}^{\rm perc} (\xi_{i}^{\rm perc})^{4}, \nonumber \\
    g_{*s,\rm tot}^{\rm perc} & \simeq g_{*s, \rm SM}^{\rm perc} + g_{*s,-1}^{\rm perc} (\xi_{-1}^{\rm perc})^{3} + \sum_{i>0} g_{*s,i}^{\rm perc} (\xi_{i}^{\rm perc})^{3},   \nonumber \\
    g_{*s,\rm tot}^{0} & \simeq g_{*s, \rm SM}^{0} + g_{*s,-1}^{0} (\xi_{-1}^{0})^{3} + \sum_{i>0} g_{*s,i}^{0} (\xi_{i}^{0})^{3}.
\end{align}

We now return to discuss our assumptions regarding nature of the exotic sector QCD phase transition as well as the key parameters governing the GW spectrum, namely, $\alpha_{-1}$, $\beta/H$, $v_{\rm w}$, and the efficiency factors. 
In principle, if the temperature-dependent effective potential describing the phase transition is known, one can compute these quantities.
However, in our scenario, due to the associated strong dynamics, we are not able to provide a first principles analysis of the phase transition properties.
Ideally, the phase transition could be studied using lattice methods (see Ref.~\cite{Aarts:2023vsf} for some perspectives), though there are no existing studies which map on to our scenario. Attempts have been made in literature to model the effective potential and resulting GW signal in certain strongly-coupled QCD-like gauge theories using various phenomenological approaches/toy models, including linear sigma models, Polyakov-Nambu-Jona-Lasino models, and holographic models, see Refs.~\cite{Bai:2018dxf,Helmboldt:2019pan,Bigazzi:2020avc,Halverson:2020xpg,Huang:2020crf,Reichert:2021cvs} for some recent representative studies. 
In many cases, these studies indicate relatively small (large) values of the phase transition strength (duration) parameters. For the Nnaturalness model, while it is not guaranteed, such values may still be potentially detectable by future space-based GW observatories, as we will discuss in Sec.~\ref{sec:res}. 
We will not attempt to model the effective potential in this work, but will instead remain agnostic about the evolution of the phase transition. To illustrate the range of possibilities, we will consider several representative benchmark scenarios for the behavior of the phase transition and the parameters governing the spectrum, as we explain in the following.

Once a bubble nucleates the bubble wall experiences negative pressure from the potential difference between the true and false vacua causing it to accelerate. At the same time, the bubble wall faces pressure from the plasma in the symmetric phase which acts as friction on the expanding bubble wall. The largest frictional pressure a bubble wall faces is when the wall velocity $v_{\rm w}$ approaches the Jouguet velocity $v_{J}$ \cite{Laurent:2022jrs}. If the negative pressure from the potential difference between the true and false vacua $(\Delta V)$ is large enough (corresponding to large $\alpha_{-1}$) to overcome this maximum pressure from the plasma, the walls will exhibit ultra-relativistic velocities corresponding to a runaway scenario. On the other hand if the $\Delta V$ is not large enough (corresponding to a smaller $\alpha_{-1}$) to overcome the maximum frictional pressure from plasma,
the wall reaches a terminal velocity with $v_{\rm w} \sim c_{\rm s}$ corresponding to a non-runaway scenario.
Following the analysis in~\cite{Ai:2023see}, we have checked that the boundary on the strength parameter between the runaway and non-runaway scenarios is given by $\alpha_{-1} \approx 0.3$ with larger $\alpha_{-1}$ corresponding to a runaway wall and smaller $\alpha_{-1}$ corresponding to a non-runaway wall with a terminal wall velocity that can be approximated by the speed of sound in the plasma, $v_{\rm w} \approx c_{\rm s} = 1/\sqrt{3}$.

Another important parameter governing the GW spectrum is $\beta/H$, which is defined as
\begin{equation}
\label{eq:betaoH}
\frac{\beta}{H}  = T_{-1} \frac{d}{dT_{-1}}  \frac{S_3}{T_{-1}}   \bigg\vert_{T_{-1}^{\rm nuc}},
\end{equation}
where $S_3$ is the three-dimensional Euclidean bounce action, assuming the phase transition proceeds due to thermal fluctuations. The parameter $\beta$ gives a measure of the duration of the phase transition.
As is clear from Eq.~\eqref{eq:betaoH}, the calculation of $\beta/H$ requires knowledge the tunneling action $S_3$ and hence the thermal potential during the phase transition, which, as mentioned above, is obscure due to the QCD strong dynamics. 
We will thus consider several benchmark choices for the phase transition duration parameter in the broad range $\beta/H \in [3,10^{4}]$. 
The lower bound is imposed to ensure efficient bubble percolation~\cite{Freese:2022qrl}.  
We note that the duration parameter $\beta/H$ is expected to be inversely correlated with the strength parameter $\alpha_{-1}$, see, e.g., Ref.~\cite{Ellis:2020awk} for discussion.

For our runaway 
scenario the energy of the phase transition is converted into  accelerating the bubble wall implying that the dominant source of GWs is bubble collisions. 
For our non-runaway 
scenario with a terminal wall velocity, the expanding wall pushes on the plasma in the symmetric phase, creating a coherent motion of the plasma. Therefore, most of the latent heat released during the phase transition is converted to sound waves. Given the above considerations, we will study the following two scenarios, which are illustrative of the range possibilities for the properties of the phase transition:
\begin{itemize}
 \setlength\itemsep{-1em}
    \item Runaway scenario:
    \vspace{-5pt}
    \begin{align}
 v_{\rm w} = 1, ~~ \kappa_{\rm BW} = 1, ~~~ \kappa_{\rm SW} = 0, ~~~~~ \nonumber \\
 (\alpha_{-1},\beta/H) = (10,3), ~(5,10), ~(1, 10^3). 
 \label{eq:runaway}
    \end{align}
    \item Non-runaway scenario:
        \begin{align}
 v_{\rm w} = \frac{1}{\sqrt{3}}, ~~ \kappa_{\rm BW} = 0, ~~~ \kappa_{\rm SW} = \frac{\alpha^{2/5}_{-1}}{0.017 +(0.997+\alpha_{-1})^{2/5}}, \nonumber \\
 (\alpha_{-1},\beta/H) = (0.3,10^2), ~(0.1,10^3), ~(0.05, 10^4).   ~~~~~~~
   \label{eq:nonrunaway}
  \end{align}
\end{itemize}   
For the non-runaway scenario with $v_{\rm w} = 1/\sqrt{3}$, we have used the numerical fitting function for the efficiency factor $\kappa_{\rm SW}$ from Ref.~\cite{Espinosa:2010hh}

The amplitude of the GW signal is governed by the strength parameter $\alpha_{\rm tot}$ in Eq.~\eqref{eq:alphas}, which is given by the product of $\alpha_{-1}$ and the fraction of the total energy density stored in the $i = -1$ sector.
It is useful to ask how large this parameter may be while maintaining consistency with the $\Delta N_{\rm eff}$ constraints discussed in the previous section. Focusing on the regions of parameter space in which the first exotic sector provides the dominant contribution to additional relativistic degrees of freedom, $\Delta N_{\rm eff} \approx \Delta N_{\rm eff,-1}$, we may then relate $\alpha_{\rm tot}$ to $\Delta N_{\rm eff,-1}$ as follows: 
\begin{align}
    \label{eq:alphatot}
    \alpha_{\rm tot} &\simeq \alpha_{-1} \left[\frac{g_{*\rho,-1}^{\rm perc}}{g_{*\rho,\rm SM}^{\rm perc}}(\xi_{-1}^{\rm perc})^4\right], \nonumber\\
    & = \frac{7}{8}\left( \frac{4}{11}  \right)^{4/3}  \frac{\alpha_{-1}}{1+ \alpha_{-1}} 
     \left[   \frac{g_{*s,-1}^{\rm CMB}  }{ g_{*s,{\rm SM}}^{\rm CMB}  }  \frac{g_{*s,{\rm SM}}^{\rm rh}  }{ g_{*s,-1}^{\rm rh}  }   \right]^{4/3} 
      \left[   \frac{g_{*\rho,-1}^{\rm rh}  }{ g_{*\rho,{\rm SM}}^{\rm perc}  }    \frac{2  }{ g_{*\rho,-1}^{\rm CMB}  }  \right]  \Delta N^{\rm CMB}_{\rm eff,-1},
     \nonumber\\ 
    & \simeq \begin{cases}
     0.02 \times \displaystyle{ \left(\frac{\Delta N_{\rm eff,-1}^{\rm CMB}}{0.7}\right)}~~~~ (\alpha_{-1}=0.3,~~{\text{Non-runaway~scenario}}),\\
     0.1 \times  \displaystyle{ \left(\frac{\Delta N_{\rm eff,-1}^{\rm CMB}}{0.7}\right)}~~~~~ (\alpha_{-1} = 10,~~{\rm Runaway~scenario}).
    \end{cases}
\end{align}
The first line follows from Eq.~\eqref{eq:alphas} given that $\rho^{\rm perc}_{\rm tot} \approx \rho^{\rm perc}_{\rm SM}$ for parameters consistent with $\Delta N_{\rm eff}$ bounds. 
In the second line we have first related $\xi_{-1}^{\rm perc}$ to $\xi_{-1}^{\rm rh}$ using Eq.~\eqref{eq:Trhi}, and then related $\xi_{-1}^{\rm rh}$ to $\Delta N_{\rm eff,-1}^{\rm CMB}$ using entropy conservation and Eq.~\eqref{eq:DNeff1}, assuming $\Delta N_{\rm eff} \approx \Delta N_{\rm eff,-1}$. 
  Eq.~\eqref{eq:alphatot}
 demonstrates that the strength parameter $\alpha_{\rm tot}$ may potentially be large enough to enable a detectable GW signal while satisfying bounds on additional relativistic degrees of freedom. 

\section{Results and Discussion}
\label{sec:res}

\begin{figure*}
\includegraphics[width=0.9\linewidth]{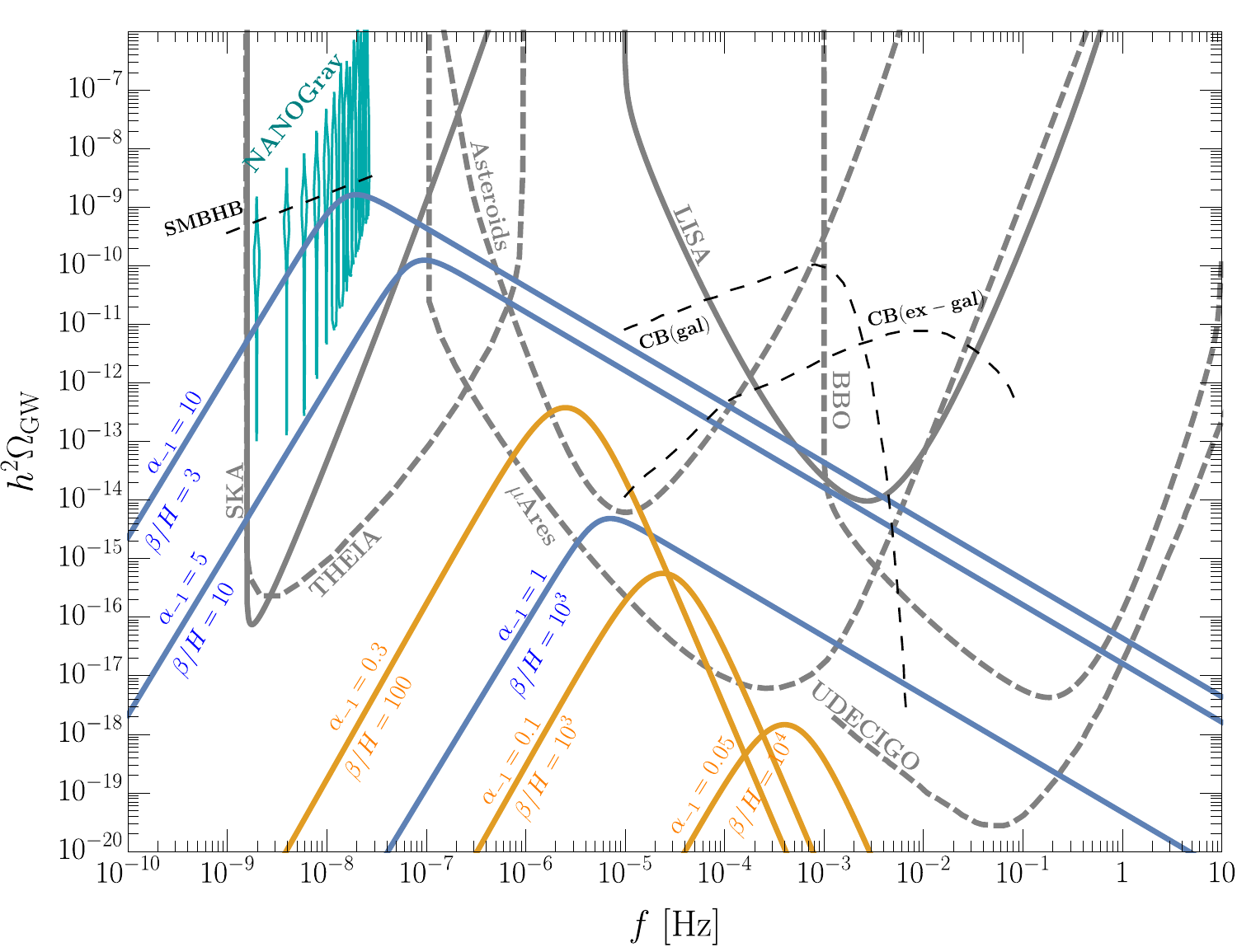}
\caption{The GW spectrum from the runaway (blue) and non-runaway (orange) scenarios defined in 
Eqs.~\eqref{eq:runaway} and~\eqref{eq:nonrunaway}, respectively.
To assess how large the signal can be, we assume the exotic sector contributes $\Delta N_{\rm eff,-1}^{\rm CMB} = 0.7$, saturating the bound from  Planck + Lensing + BAO + SH0ES. 
Also shown are the NANOGrav 15-yr results (teal violin), the PLIS curves for upcoming experiments SKA and LISA (solid gray) and proposed experiments THEIA, $\mu$Ares, asteroid laser ranging, and BBO (dashed gray). 
The PLIS curves for SKA, LISA, and BBO are adopted from~\cite{Schmitz:2020syl} but scaled to observation times of $20~{\rm yrs}$~\cite{Janssen:2014dka} for SKA, $3~{\rm yrs}$ for LISA~\cite{Caprini:2019pxz}, and $4~{\rm yrs}$ for BBO~\cite{Crowder:2005nr}. 
The $\mu$Ares PLIS is taken from~\cite{Sesana:2019vho}, scaled to $\rm{SNR} = 1$. 
For the asteroid ranging proposal, we adopt the strain sensitivity given in~\cite{Fedderke:2021kuy} and calculate the PLIS curve using the procedure outlined in~\cite{Caprini:2019pxz} for $\rm{SNR} = 1$ and assumed experiment duration of $7~{\rm yrs}$. 
For THEIA we adopt the PLIS sensitivity calculated in~\cite{Garcia-Bellido:2021zgu} for ${\rm SNR}=1$ and a mission lifetime of $20~{\rm yrs}$. For Ultimate-DECIGO (UDECIGO) we have adopted the PLIS in~\cite{Braglia:2021fxn}.
Black dashed lines represent foregrounds from galactic and extragalactic compact binaries (CB)~\cite{Robson:2018ifk,Farmer:2003pa} and the SMBHB best fit to the NANOGrav 15-yr measurement~\cite{NANOGrav:2023gor}. 
}
\label{fig:GWplot}
\end{figure*}

\begin{figure} [htb]
 \includegraphics[width=0.43\linewidth]{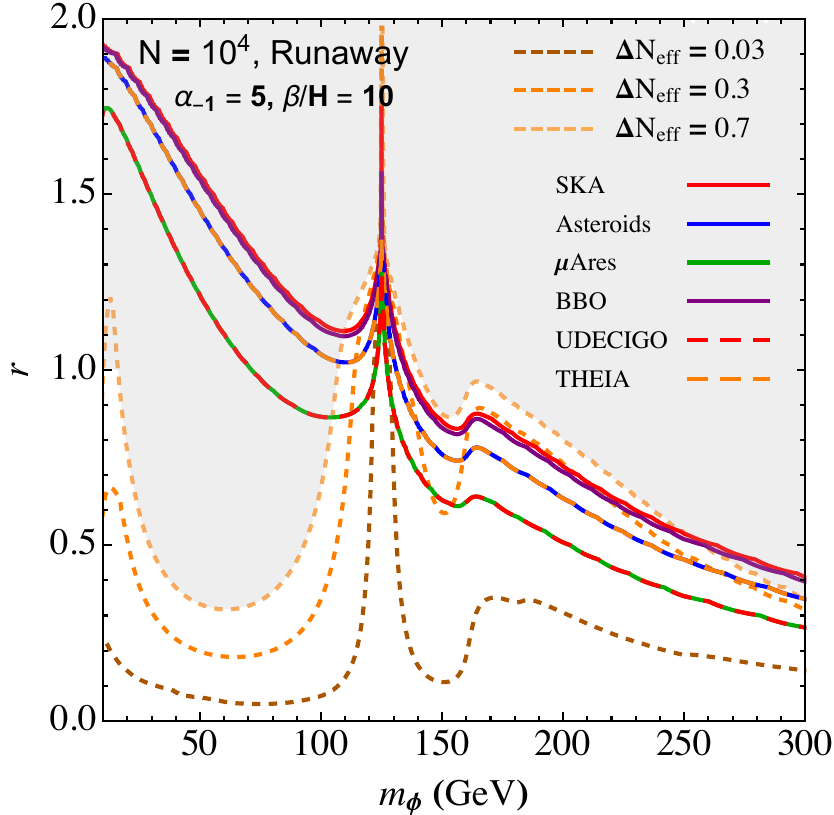} \quad\quad
 \includegraphics[width=0.43\linewidth]{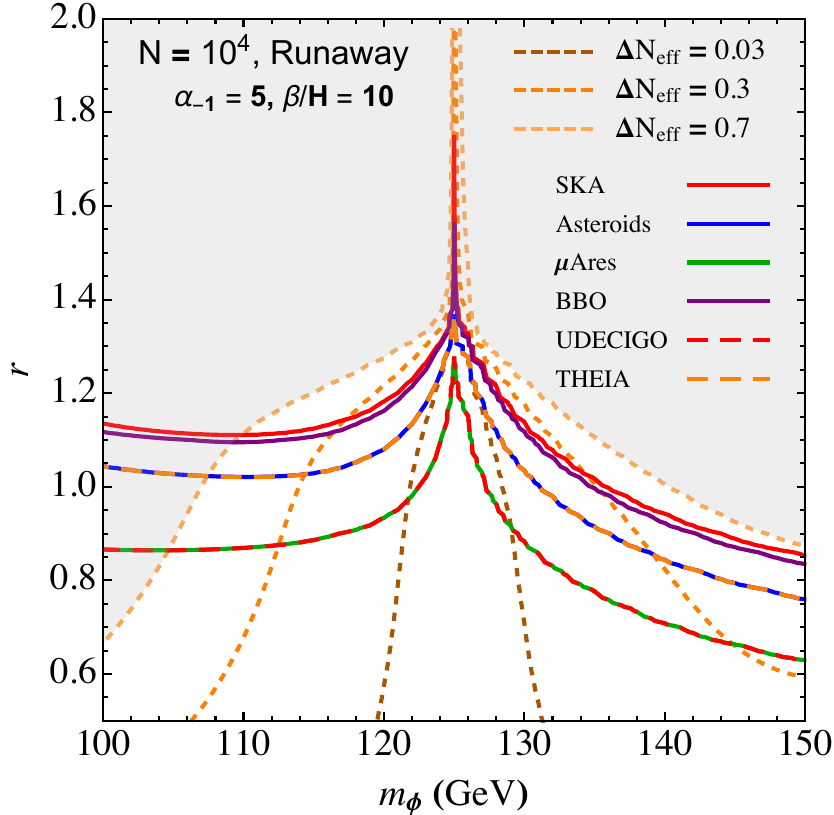} \\
 \includegraphics[width=0.43\linewidth]{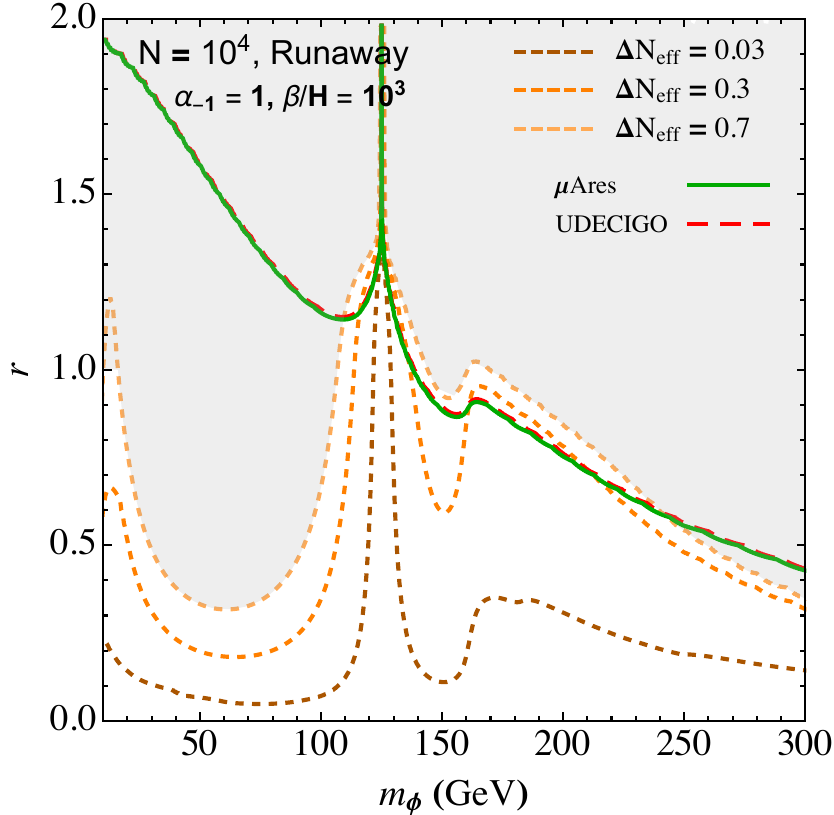}  \quad\quad
 \includegraphics[width=0.43\linewidth]{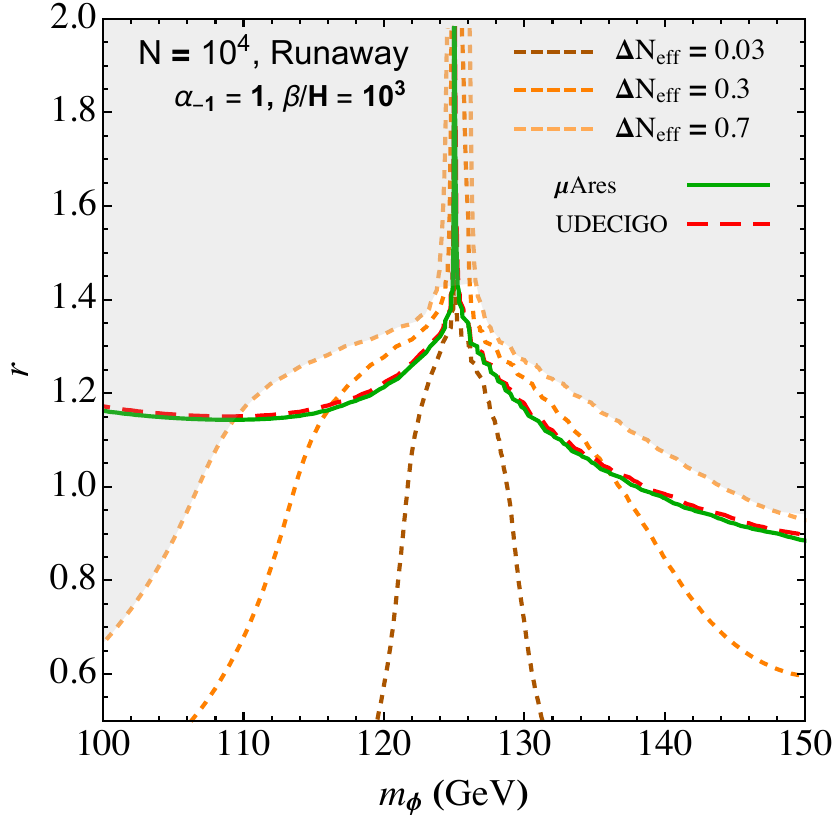} 
 \caption{Reach from various gravitational wave experiments in the $r$ vs. $m_\phi$ space over the full parameter region (left) and zoomed into the Higgs funnel (right).  Results are shown for the runaway scenario with $\beta/H=10$ and $\alpha_{-1}=5$ (top) and $\beta/H=10^3$ and $\alpha_{-1}=1$ (bottom).  Short dashed lines indicate contours of $\Delta N_{\rm eff}$ evaluated at $T^{\rm CMB}$.}
 \label{fig:parameterSpaceRunaway}
\end{figure}

\begin{figure} [htb]
 \includegraphics[width=0.43\linewidth]{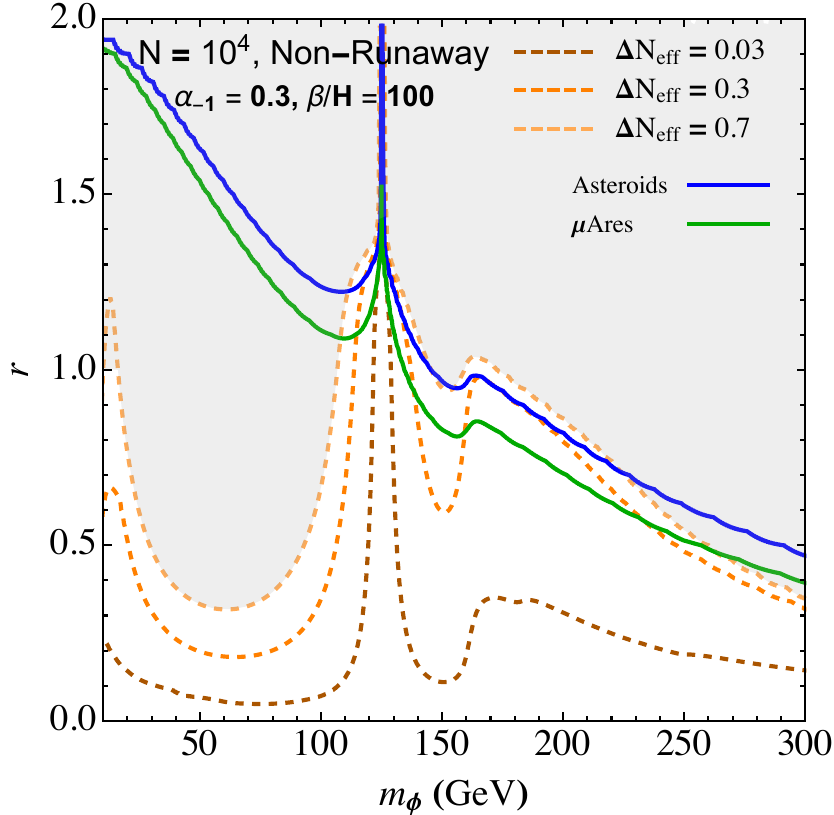}  \quad\quad
 \includegraphics[width=0.43\linewidth]{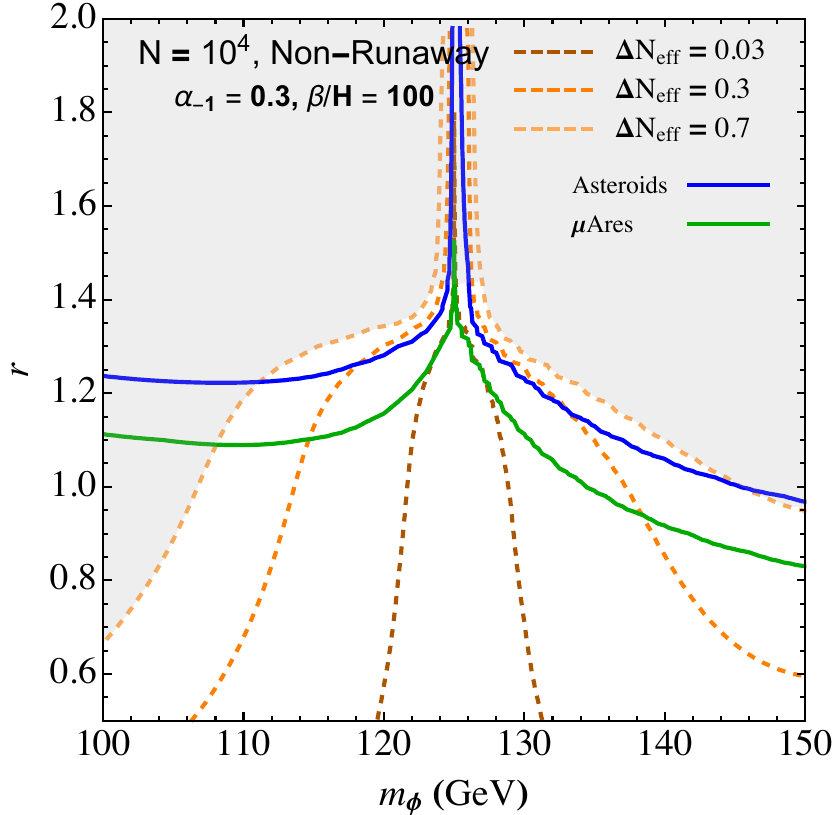}  \\
 \includegraphics[width=0.43\linewidth]{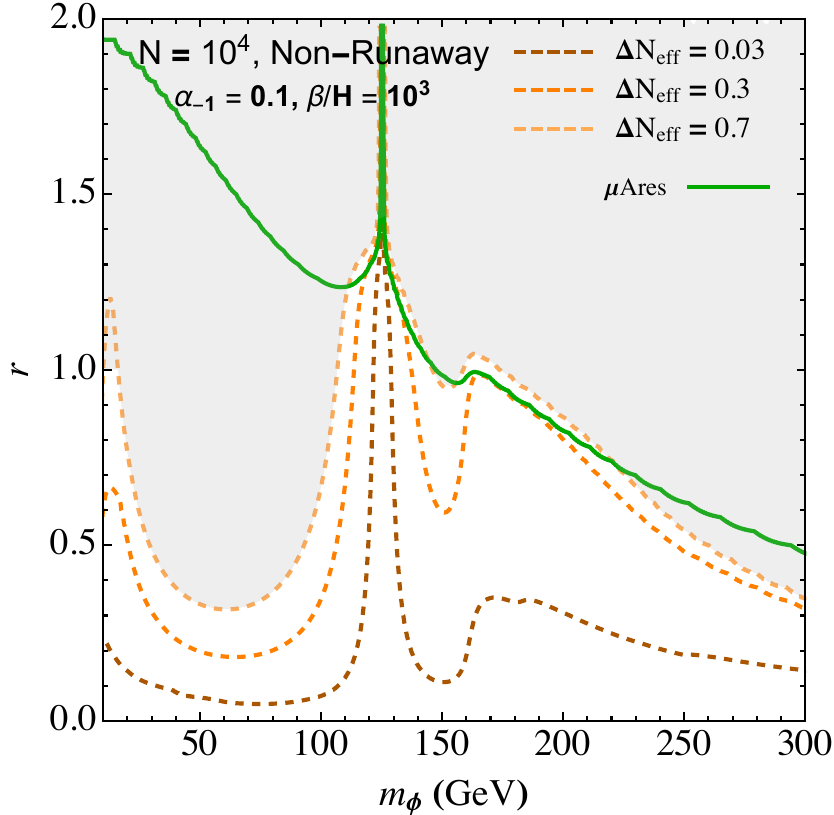}  \quad\quad
 \includegraphics[width=0.43\linewidth]{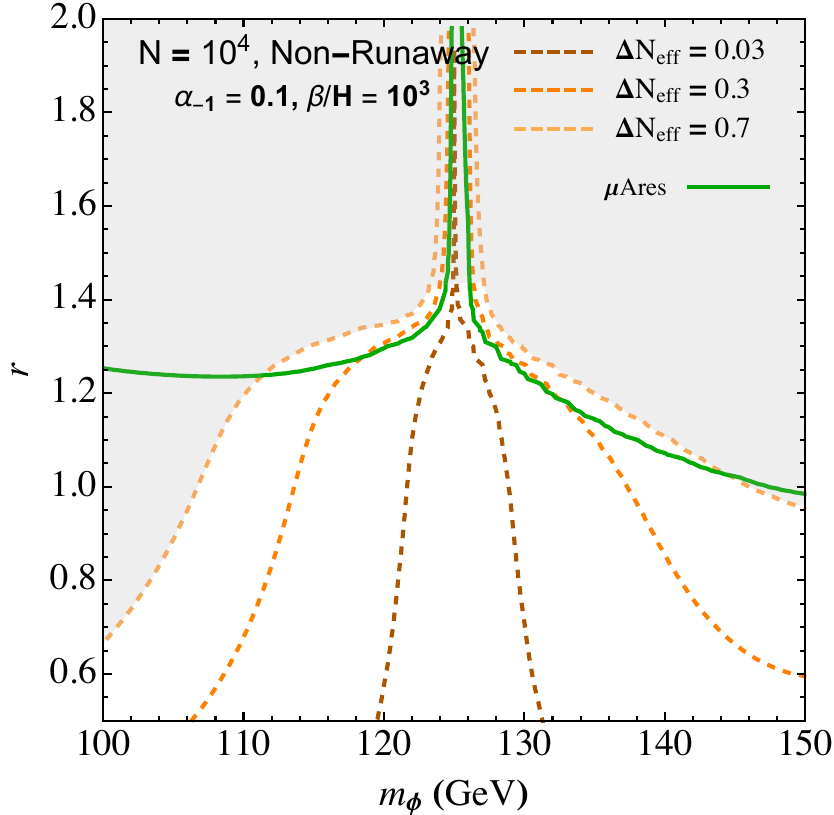} 
 \caption{Reach from various gravitational wave experiments in the $r$ vs. $m_\phi$ space over the full parameter region (left) and zoomed into the Higgs funnel (right).  Results are shown for the non-runaway scenario with $\beta/H=100$ and $\alpha_{-1}=0.3$ (top) and $\beta/H=10^3$ and $\alpha_{-1}=0.1$ (bottom).  Short dashed lines indicate contours of $\Delta N_{\rm eff}$ evaluated at $T^{\rm CMB}$.}
 \label{fig:parameterSpaceNonrunaway}
\end{figure}

Using the results of the previous section, in Fig.~\ref{fig:GWplot} we show the GW spectrum for the runaway and non-runaway scenarios defined in 
Eqs.~\eqref{eq:runaway} and~\eqref{eq:nonrunaway}, respectively.  To exhibit the maximal allowed strength of the GW signal, we have saturated the Planck + Lensing + BAO + SH0ES bound on additional relativistic species, taking  $\Delta N_{\rm eff,-1}^{\rm CMB} = 0.7$.\footnote{ $\Delta N_{\rm eff,-1}^{\rm CMB} = 0.7$ is only used for Fig.~\ref{fig:GWplot}.  When results are shown in the Nnaturalness parameter space, as in Figs.~\ref{fig:parameterSpaceRunaway} and~\ref{fig:parameterSpaceNonrunaway} the $\Delta N_{\rm eff}$ limits are compared to $\Delta N_{\rm eff,tot}^{\rm CMB}$.}
We compare our predictions to the power law integrated sensitivity (PLIS) curves corresponding to a signal-to-noise ratio (SNR) threshold of 1 for several future GW experiments, 
including the SKA PTA~\cite{Janssen:2014dka};
the spaced-based interferometers LISA~\cite{2017arXiv170200786A}, BBO~\cite{Crowder:2005nr}, Ultimate-DECIGO \cite{Braglia:2021fxn}, $\mu$Ares~\cite{Sesana:2019vho}, and asteroid laser ranging~\cite{Fedderke:2021kuy}; and future astrometric measurements \cite{Moore:2017ity} for the proposed THEIA experiment~\cite{2018FrASS...5...11V}.
We also show the  stochastic GW spectrum from the NANOGrav 15-yr result~\cite{NANOGrav:2023hvm}.
Finally, estimates for 
astrophysical foregrounds coming from supermassive black hole binaries (SMBHBs)~\cite{NANOGrav:2023gor}, as well as galactic~\cite{Cornish:2017vip} and extragalactic compact binaries~\cite{Farmer:2003pa}, are also displayed in Fig.~\ref{fig:GWplot}.

Assuming that the astrophysical foregrounds either can be resolved and subtracted (see Ref.~\cite{2012PhRvD..85d4034B} for SMBHB foreground resolution) or are somewhat weaker in strength than currently expected,  Fig.~\ref{fig:GWplot} demonstrates that there are promising opportunities to probe Nnaturalness with future GW measurements. As emphasized several times, this depends sensitively on the precise nature of the first exotic sector QCD phase transition, about which there are significant theoretical uncertainties, as well as the fractional energy density contained in this sector, which is dictated by the Nnaturalness model parameters, as we will discuss in detail shortly. Runaway transitions can be probed by PTAs, astrometric measurements, and spaced-based interferometers, while non-runaway transitions could lead to a signal in space-based interferometers. 
It is also clear from Fig.~\ref{fig:GWplot} that even under the most optimistic assumptions (runaway transitions, large $\alpha_{-1}$, small $\beta/H$, and maximal $\Delta N_{\rm eff,-1}^{\rm CMB}$) it is unlikely that Nnaturalness can fully account for the stochastic gravitational background reported in the NANOGrav 15-yr dataset (a similar point was made recently for the NANOGrav 12.5-yr dataset for generic stable secluded sectors~\cite{Bringmann:2023opz}). 
Furthermore, we observe that if $\alpha_{-1}$ is too small, or $\beta/H$ is too large, as may be suggested by detailed studies of FOPTs in toy models of QCD-like theories (see discussion in previous section), the GW signal from Nnaturalness may lie outside the reach of proposed experiments. The different scenarios considered here, Eqs.~(\ref{eq:runaway},\ref{eq:nonrunaway}), serve to illustrate the range of possibilities. 

Next, we map out the regions of the Nnaturalness parameter space that can potentially be probed by future GW experiments. Specifically, we determine the values of  $m_\phi$ and $r$ that yield a GW signal intersecting (or tangential to) the PLIS curves shown in Fig. \ref{fig:GWplot}. 
In Fig.~\ref{fig:parameterSpaceRunaway} we show this reach for two runaway scenarios, 
$(\alpha_{-1},\beta/H) = ~(5,10)$ (top) and  $(\alpha_{-1},\beta/H) = ~(1, 10^3)$ (bottom). For each scenario, we show both the full parameter space for a reheaton mass lighter than 300 GeV (left), as well as a zoomed-in region of parameter space near $m_\phi \approx m_h$ (right).\footnote{We note that for the non-runaway scenarios we have made the conservative choice of not considering the contribution to GWs from turbulence in the plasma due to the associated theoretical uncertainties. Including that contribution will improve the observational reach for these scenarios.} 
The figures also show the predictions for $\Delta N_{\rm eff}^{\rm CMB}$ and the region excluded by the Planck + Lensing + BAO + SH0ES data. 
In the most optimistic scenario (Fig.~\ref{fig:parameterSpaceRunaway}, top), there are several experiments and techniques (PTAs, space-based interferometry, astrometry) that can explore uncharted Nnaturalness parameter space, and, in particular, $\mu$Ares, Ultimate-DECIGO, and THEIA even have the potential to compete in reach with future precision CMB measurements, e.g., CMB Stage IV~\cite{Baumann:2015rya} ($\Delta N_{\rm eff}^{\rm CMB} \lesssim 0.03$). 

Similarly, in Fig.~\ref{fig:parameterSpaceNonrunaway} we show the reach of future GW experiments for two non-runaway scenarios, 
$(\alpha_{-1},\beta/H) = ~(0.3,10^2)$ (top) and  $(\alpha_{-1},\beta/H) = ~(0.1, 10^3)$ (bottom). For these scenarios, the GW signal from sound waves is predicted to lie in the $\mu$Hz range and thus can potentially be probed by future space-based interferometers such as $\mu$Ares and asteroid ranging.
It is worth noting that the phase transition parameters for the second scenario, $(\alpha_{-1},\beta/H) = ~(0.1, 10^3)$ (bottom), are broadly consistent with results from studies modeling the phase transitions of QCD-like theories.

The behavior of the GW sensitivity curves in Figs.~\ref{fig:parameterSpaceRunaway} and~\ref{fig:parameterSpaceNonrunaway} can be understood by recalling that the strength parameter $\alpha_{\rm tot}$ is approximately linearly related to $\Delta N_{\rm eff, -1}$ in the parameter regions where the total energy density is dominated by the SM bath, see Eq.~\eqref{eq:alphatot}. Thus, the GW sensitivity curves largely overlap with isocontours of $\Delta N_{{\rm eff},-1}$, as can be seen by comparing with Fig.~\ref{fig:mPhi_r_parameterSpace}. The reach of GW experiments is strongest in the regions of parameter space where the reheaton has a relatively sizable branching ratio into the first exotic sector. 
In the allowed regions of parameter space, consistent with CMB constraints on $\Delta N_{\rm eff}$, this may occur in the regions where $m_\phi \lesssim 2 m_{H_{-1}}$ and the reheaton decays dominantly to the SM. These two requirements combine to sculpt two regions where GW experiments can be sensitive: 1) near the ``Higgs funnel'', $m_\phi \approx m_h$, and 2) above the threshold for reheaton decays to SM gauge bosons, as is observed in Figs.~\ref{fig:parameterSpaceRunaway} and~\ref{fig:parameterSpaceNonrunaway}. 

\section{Conclusions}
\label{sec:con}

Nnaturalness is a novel approach to the hierarchy problem. The key prediction of the framework is the existence of many decoupled hidden sectors containing small fractional energy densities, which can be probed through cosmological measurements such as $\Delta N_{\rm eff}$. In this work, have explored the potential to probe Nnaturalness through GW observations. Considering the scalar reheaton model for concreteness, in certain parameter regions the first exotic sector, with the smallest positive squared Higgs mass, is predicted to have a sizable fractional energy density. QCD in this sector is expected to feature a cosmological first-order chiral symmetry breaking phase transition since all quarks are much lighter than the confinement scale, which then yields an associated stochastic GW signal.  The resulting GW spectra are expected to peak in the ${\rm nHz} - {\rm mHz}$ frequency range, 
with a strength scaling with the fraction of the reheaton energy density stored in the first exotic sector. 

We have delineated the regions of parameter space where the first exotic sector has a substantial energy density, and thus a potentially detectable stochastic GW signature. The observational prospects for the GW signal depends sensitively on the detailed evolution of the exotic sector QCD phase transition, which involves strong coupling dynamics. We have remained agnostic about the phase transition properties, exploring several scenarios designed to encompass the spectrum of conceivable possibilities.
Depending on these assumptions, as well as the eventual capabilities to discriminate various astrophysical foregrounds, we find that a GW signal from Nnaturalness can potentially be observable in several future experiments, including PTAs (SKA), planned (BBO) and proposed (Ultimate-DECIGO, $\mu$Ares, asteroid ranging) spaced-based interferometers, and astrometric measurements (THEIA). In some of the more optimistic phase transition scenarios, future GW observations may even complement tests of Nnaturalness from next generation CMB experiments such as CMB Stage IV. 

Our study reveals several interesting open questions. 
First, it would be valuable to further clarify the nature of the exotic sector QCD phase transition, using lattice studies as well as phenomenological models. 
Additional uncertainties in our predictions come from the modeling of the GW production from a first-order phase transition (see, e.g., Ref.~\cite{Athron:2023rfq} for a recent discussion). 
Progress on these issues will lead to a better understanding of the capabilities of GW observatories to probe Nnaturalness.

We have only considered the scalar reheaton model in this work, and it would be very interesting to also explore the potential gravitational wave signatures of the other reheaton models considered in Ref.~\cite{Arkani-Hamed:2016rle}. More accurate studies of the cosmological perturbations and their impact on the CMB and structure formation in the regions where the exotic sectors are populated would also be valuable (for a study considering the SM-like sectors, see Ref~\cite{Bansal}). The exotic sector QCD FOPT phase transition could also be associated with other novel phenomena, such as the formation of dark quark nuggets~\cite{Witten:1984rs,Bai:2018dxf} in the presence of a corresponding baryon asymmetry, or the production of primordial black holes~\cite{Hawking:1971ei,Carr:1974nx,Crawford:1982yz}. 
Future studies along these directions may point the way to even more new probes of Nnaturalness.

\section{Acknowledgements}
We thank  Amit Bhoonah, Raffaele Tito D'Agnolo, Arnab Dasgupta, Michael Fedderke, Subhajit Ghosh, Arthur Kosowsky, Pedro Schwaller, Bibhushan Shakya, Qian Song, and Yuhsin Tsai for helpful discussions and correspondence. 
The work of B.B., A.G., M.L., and M.R. is supported by the U.S. Department of Energy under grant No. DE–SC0007914.  
M.L. is also supported by the National Science Foundation under grant no. PHY-2112829.
M.R. is also supported by the U.S. Department of Energy under grant No. DE-SC0010813.

\appendix

\section{Nnaturalness Decay Widths}
\label{sec:nn}

\subsection*{Standard Model-Like Sectors}

For a SM-like sector $i$, the reheaton decays via mixing with the Higgs $h_i$ from that sector to all final states $\{ f_i \}$ that are kinematically open
\begin{equation}
\Gamma_{\phi \to \{ f_i \} } = \theta_i^2 \; \Gamma_{h_i \to \{ f_i \} }(m_\phi),
\end{equation}
where $\theta_i$ is the mixing between $\phi$ and $h_i$.

When the reheaton is heavy relative to $h_i$ the decay $\phi \to h_i h_i$ can open
\begin{equation}
\Gamma_{\phi \to h_i h_i} = \frac{a^2}{32\pi m_\phi} \sqrt{1-\frac{4m_{h_i}^2}{m_\phi^2}}.
\end{equation}

\subsection*{Exotic Sectors}

For an exotic sector $i$, electroweak symmetry is broken near $\Lambda_{\rm QCD}^{(i)}$, therefore provided that $m_\phi > \Lambda_{\rm QCD}^{(i)}$ decay widths are calculated in the unbroken electroweak phase where all vectors and fermions and massless.

There are always one-loop decays into pairs of vectors $B_{i} B_{i}$ and $W_{i}^a W_{i}^a$ ($a=1,2,3$)
\begin{equation}
\Gamma_{\phi \to B_{i} B_{i}} = 
\frac{g'^4 a^2}{4096 \pi^5 m_\phi}
|\tau A_0\left(\tau\right)|^2,
\quad\quad\quad
\tau = \frac{m_\phi^2}{4m_{H_i}^2},
\end{equation}
\begin{equation}
\Gamma_{\phi \to W_{i}^a W_{i}^a} = 
\frac{3 g^4 a^2}{4096 \pi^5 m_\phi}
|\tau A_0\left(\tau\right)|^2,
\quad\quad\quad
\tau = \frac{m_\phi^2}{4m_{H_i}^2},
\end{equation}
where $A_0(\tau)$ is given by
\begin{equation}
A_0(\tau) = \tau^{-2} (f(\tau) - \tau),
\end{equation}
\begin{equation}
f(\tau) = \begin{cases}
{\rm arcsin}^2(\sqrt{\tau})  & \tau \leq 1, \\
-\frac{1}{4}\left(\log\left(\frac{1+\sqrt{1-\tau^{-1}}}{1-\sqrt{1-\tau^{-1}}}\right)- i\pi\right)^2 & \tau > 1.
\end{cases}
\end{equation}
The diagrams that would lead to $\phi \to \bar{t}_{R,i} t_{R,i}$ and $\phi \to \bar{Q}_{L,i} Q_{L,i}$ are zero in the massless fermion limit due to helicity conservation.

When $m_\phi > 2 m_{H_{i}}$ the reheaton can decay to two on-shell Higgs particles
\begin{equation}
\Gamma_{\phi \to H_{i} H_{i}^\dagger} = \frac{a^2}{8\pi m_\phi} \sqrt{1-\frac{4m_{H_{i}}^2}{m_\phi^2}}.
\end{equation}
When $m_{H_{i}} < m_\phi < 2 m_{H_{i}}$ the three-body decay where either $H_{i}$ or $H_{i}^\dagger$ is off-shell occurs.  The leading three-body final states are $\phi \to H_{i} Q_{L,i} \bar{t}_{R,i}$ and $\phi \to H^\dagger_{i} \bar{Q}_{L,i} t_{R,i}$
\begin{equation}
\Gamma_{\phi \to H_{i} Q_{L,i} \bar{t}_{R,i}} = 
\frac{3 y_t^2 a^2}{128 \pi^3 m_\phi^3}
\int_0^{(m_\phi - m_{H_{i}})^2} ds 
\frac{s \lambda^{1/2}(m_\phi^2,m_{H_{i}}^2,s)}{(s-m_{H_{i}}^2)^2 + (m_{H_{i}} \Gamma_H)^2},
\end{equation}
where $y_t$ is the top Yukawa and 
$\lambda(x,y,z) = x^2 + y^2 + z^2 - 2xy - 2yz - 2zx$.    
Since $y_t \gg y_f$ for all other fermions $f$, the top quark dominates the three-body width.  For $m_\phi < m_{H_{i}}$, the four-body decay where both $H_{i}$ and $H_{i}^\dagger$ are off-shell occurs.

\section{Effective Relativistic Degrees of Freedom}
\label{sec:gstar}

The determination of $\Delta N_{\rm eff}^{\rm CMB}$ and $\Omega_{\rm GW}$ in the Nnaturalness model requires calculations of the number of effective relativistic degrees of freedom in each sector at several points in the cosmological history. For our results presented in the main text, we numerically determine $g_{*\rho,{i}}$ and $g_{*s,{i}}$ for each sector $i$ based on the spectrum of the sector and the temperature of the sector at the relevant epoch, which are governed by $T^{\rm RH}$ and the Nnaturalness model parameters $m_\phi$ and $r$.

In Table \ref{tab:gstar} we compile estimates for the typical values for $g_{*\rho}$ and $g_{*s}$ for the SM, first SM-like sector, and first exotic sector at the epochs of reheating from reheaton decay (RH), exotic sector QCD phase transition at the point of percolation (perc) and reheating (rh), SM sector recombination (CMB), and today (0). These values are characteristic of regions of parameter space that are both cosmologically viable and feature a relatively large fractional energy density in the first exotic sector. 

\begin{table}
\begin{center}
\begin{tabular}{|c|c|c|c||c|c|c|}
 \cline{2-7}
\multicolumn{1}{c|}{}  & \multicolumn{3}{c||}{$g_{*\rho}$}  &  \multicolumn{3}{c|}{$g_{*s}$}  \\  \cline{3-5}
\hline
 \diagbox{~Epoch~}{~Sector~} & ~SM~ & ~$i = 1$~ & ~$i = -1$~ & ~SM~ & ~$i = 1$~ & ~$i = -1$~ \\
\hline
\hline
RH &  ~$\approx 100$~ & ~$\approx 100$~ & ~$\approx 100$~ & ~$\approx 100$~ & ~$\approx 100$~ & ~$\approx 100$~ \\
\hline
perc & ~$\approx 60$~ & ~$\approx 60$~ & ~$102.75$~ & ~$\approx 60$~ & ~$\approx 60$~ & ~$\approx 102.75$~ \\
\hline
rh & ~$\approx 60$~ & ~$\approx 60$~ & ~$58.75$~ & ~$\approx 60$~ & ~$\approx 60$~ & ~$\approx 58.75$~ \\
\hline
CMB & ~$3.36$~ & ~$2$~ & ~$12.4$~ & ~$3.91$~ & ~$3.91$~ & ~$12.8$~ \\
\hline
0 & ~$2$~ & ~$2$~ & ~$7.25$~ & ~$3.91$~ & ~$3.91$~ & ~$7.81$~ \\
\hline
\end{tabular}
\end{center}
 \caption{Typical values of $g_{*\rho}$ and $g_{*s}$ for the SM, first SM-like sector, and first exotic sector for allowed parameter regions featuring a relatively large fractional energy density in the first exotic sector. Estimates are given at the epochs of reheating from reheaton decay (RH), exotic sector QCD FOPT at the point of percolation (perc) and reheating (rh), SM sector recombination (CMB), and today (0). We have assumed $T^{\rm RH} = 100$ GeV.}
\label{tab:gstar}
\end{table}

A few remarks are in order regarding Table \ref{tab:gstar}. 
First, for the initial reheat temperature we have assumed $T^{\rm RH} = 100$ GeV, leading to the estimate of roughly 100 relativistic degrees of freedom in each sector. Note that these are estimates for the purposes of Table~\ref{tab:gstar} but are calculated numerically in the results presented in the main text.
During the exotic sector QCD phase transition, we estimate $g_{*\rho,{-1}}^{\rm perc}  = g_{*s,{-1}}^{\rm perc} \approx 102.75$ (unbroken phase including all degrees of freedom except the Higgs doublet $H_{-1}$) and $g_{*\rho,{-1}}^{\rm rh} = g_{*s,{-1}}^{\rm rh} \approx 58.75$  (broken phase including $\gamma$, $W$, $Z$, charged leptons, neutrinos, pions). 
For the SM and SM-like sectors, the relativistic degrees of freedom are typically varying rapidly with temperature near the exotic sector QCD phase transition, with our choice of $60$ relativistic degrees of freedom in Table~\ref{tab:gstar} being a representative value. 
Near recombination, we have as usual $g_{*\rho,{\rm SM}}^{\rm CMB}  = 3.36$ and $g_{*s,{\rm SM}}^{\rm CMB}  = 3.91$. Furthermore, under the simplifying assumption of degenerate Dirac neutrinos in the SM-like sectors, the neutrinos in the these sectors are typically non-relativistic near recombination, leading to $g_{*\rho,i}^{\rm CMB}  = 2$ and $g_{*s,i}^{\rm CMB}  = 3.91$ for $i >0$. In the first exotic sector, accounting for the fact that neutrino decoupling occurs before all charged leptons annihilate and that photons, neutrinos, electrons, and muons are typically all relativistic near recombination, and using Eq.~(\ref{eq:T-1nu-late}), we arrive at the estimates $g_{*\rho,-1}^{\rm CMB} = 12.4$ and $g_{*s,-1}^{\rm CMB} = 12.8$. At late times, the exotic sector muons leave the bath, yielding $g_{*\rho,-1}^{0} = 7.25$ and $g_{*s,-1}^{0} = 7.81$. 

\newpage
\bibliographystyle{apsrev4-1}
\bibliography{refs}

\end{document}